\documentclass[11pt]{article}
\usepackage{amsmath,amssymb,amsthm,epsfig}
\usepackage[margin=1in]{geometry}
\usepackage{eucal} 
\usepackage{graphicx}
\usepackage{hyperref,aliascnt}
\usepackage{cite}
\usepackage{microtype}

\newtheorem{theorem}{Theorem}[section]

\newaliascnt{lemma}{theorem}
\newtheorem{lemma}[lemma]{Lemma}
\aliascntresetthe{lemma}

\newaliascnt{corollary}{theorem}
\newtheorem{corollary}[corollary]{Corollary}
\aliascntresetthe{corollary}

\newaliascnt{observation}{theorem}
\newtheorem{observation}[observation]{Observation}
\aliascntresetthe{observation}

\theoremstyle{definition}
\newaliascnt{definition}{theorem}
\newtheorem{definition}[definition]{Definition}
\aliascntresetthe{definition}

\newcommand{\derivedgraph}{\operatorname{BCS}}
\newcommand{\conflictgraph}{\operatorname{CG}}

\renewcommand{\epsilon}{\varepsilon}

\newcommand{\comment}[1]{}

\newcommand{\FPlanar}[1]{\textsc{Planar}}

\title{Finding Maximal Sets of Laminar 3-Separators\\ in   Planar Graphs in Linear Time} 
\author{David Eppstein\,\thanks{Computer Science Department, University of California, Irvine. Supported in part by NSF grants CCF-1228639, CCF-1618301, and CCF-1616248.}~\footnotemark[3]
\and Bruce Reed\,\thanks{School of Computer Science, McGill University}~\thanks{This research was initiated at the Third Annual Workshop on Geometry and Graphs, Bellairs Research Institute, Holetown, Barbados, March 2014.} }
\begin{document}
\maketitle  
\begin{abstract}
We consider decomposing a 3-connected planar graph  $G$ using laminar separators of size three. 
We show how to find a maximal set of laminar 3-separators in such a graph  in linear time. 
We also discuss how to find maximal laminar set  of  3-separators from special families. 
For example we discuss non-trivial cuts, ie. cuts  which split $G$ into two components of size at least two.  
For  any vertex $v$, we also show how to find a maximal set of  3-separators  disjoint from $v$ which are laminar 
and satisfy: every vertex in a separator  $X$ has two neighbours  not in the unique    component of $G-X$  containing $v$. 
In all  cases, we show   how to construct a corresponding tree decomposition of adhesion three. Our new algorithms form an important component of recent methods for finding disjoint paths in nonplanar graphs.
\end{abstract}

\section{Introduction}

Beginning with the work of Wagner~\cite{Wag-MA-37}, decompositions of graphs by laminar sets of small cutsets have been an important tool in graph structure theory and (later) in graph algorithms. 
A subset~$X$ of the vertices of a graph $G$ forms a \emph{cutset} if $G-X$ is disconnected. It is a \emph{trivial cutset} if $G-X$ has only two components, one of which is a single vertex, and a \emph{nontrivial cutset} otherwise. A family of cutsets is \emph{laminar} if, for every two cutsets $X$ and $X'$ in the family, there exists a component $Y$ of $G-X$ such that $X'\subset X\cup Y$. Such a family can be described by a tree in which the tree nodes are the cutsets and the pieces that they decompose the graph into, and the tree edges connect cutsets with adjacent pieces.
Wagner showed, for instance, that every $K_5$-minor-free graph can be decomposed by a laminar system of cutsets of size at most three into pieces that are either planar or the eight-vertex Wagner graph. Many similar decomposition theorems are now known; in particular, tree decompositions and treewidth are also defined by laminar cutsets.

In any graph, the cutsets of cardinality one (articulation vertices) are automatically laminar, and a \emph{block-cut tree} describing their laminar tree structure can be constructed in linear time~\cite{HopTar-CACM-73}.
The cutsets of cardinality two are not necessarily laminar (for instance, in a cycle graph, every non-adjacent pair of vertices is a cutset) but they can be described by a tree structure, the SPQR tree, whose nodes represent 3-connected components of the graph~\cite{Mac-DMJ-37,DiBTam-ICALP-90}. From this structure, it is easy to recover a maximal laminar family of 2-cutsets. Again, the SPQR tree and a corresponding maximal laminar family can be constructed in linear time~\cite{GutMut-GD-01}.
 And in maximal planar graphs, the cutsets of cardinality three (separating triangles) are automatically laminar, and all separating triangles can be found in linear time~\cite{ChiNis-SJC-85}. Each of these decompositions gives a natural way of partitioning a graph into pieces with greater connectivity and therefore a more constrained structure. Many algorithms for graphs and particularly for planar graphs use these decompositions to reduce the problem they solve to simpler cases.

However,
this nice picture of canonical and automatically-laminar sets of partitions, constructed in linear time, breaks down for $3$-connected planar graphs that are not maximal. In such graphs, there may exist a nonlinear number of $3$-cutsets (also called separating triples or $3$-separations) and not all $3$-cutsets are laminar. For instance, the wheel graphs have quadratically many $3$-cutsets, but only linearly many of these can form a laminar family. Therefore, it would not be efficient to list all $3$-cutsets and then select a maximal laminar subset of them. Augmenting a $3$-connected planar graph to be maximal, and listing all separating triangles in the augmented graph, may produce a  laminar set of $3$-cutsets that is not maximal in the original graph. Thus, the first unresolved case for finding laminar cutsets efficiently in planar graphs is the case when the graph is $3$-connected but not maximal and we seek to decompose it according to its $3$-cutsets. Such decompositions have been studied before, and applied as part of other algorithms for planar graphs and related graph classes, but with a focus on logarithmic space complexity or on applications of these decompositions rather than as here on the time efficiency for finding the decomposition~\cite{ThiWag-CJTCS-14,StrThiWag-TCS-16,ChaEpp-JGAA-13,EppVaz-18}.

\begin{figure}[t]
\centering\includegraphics[scale=0.4]{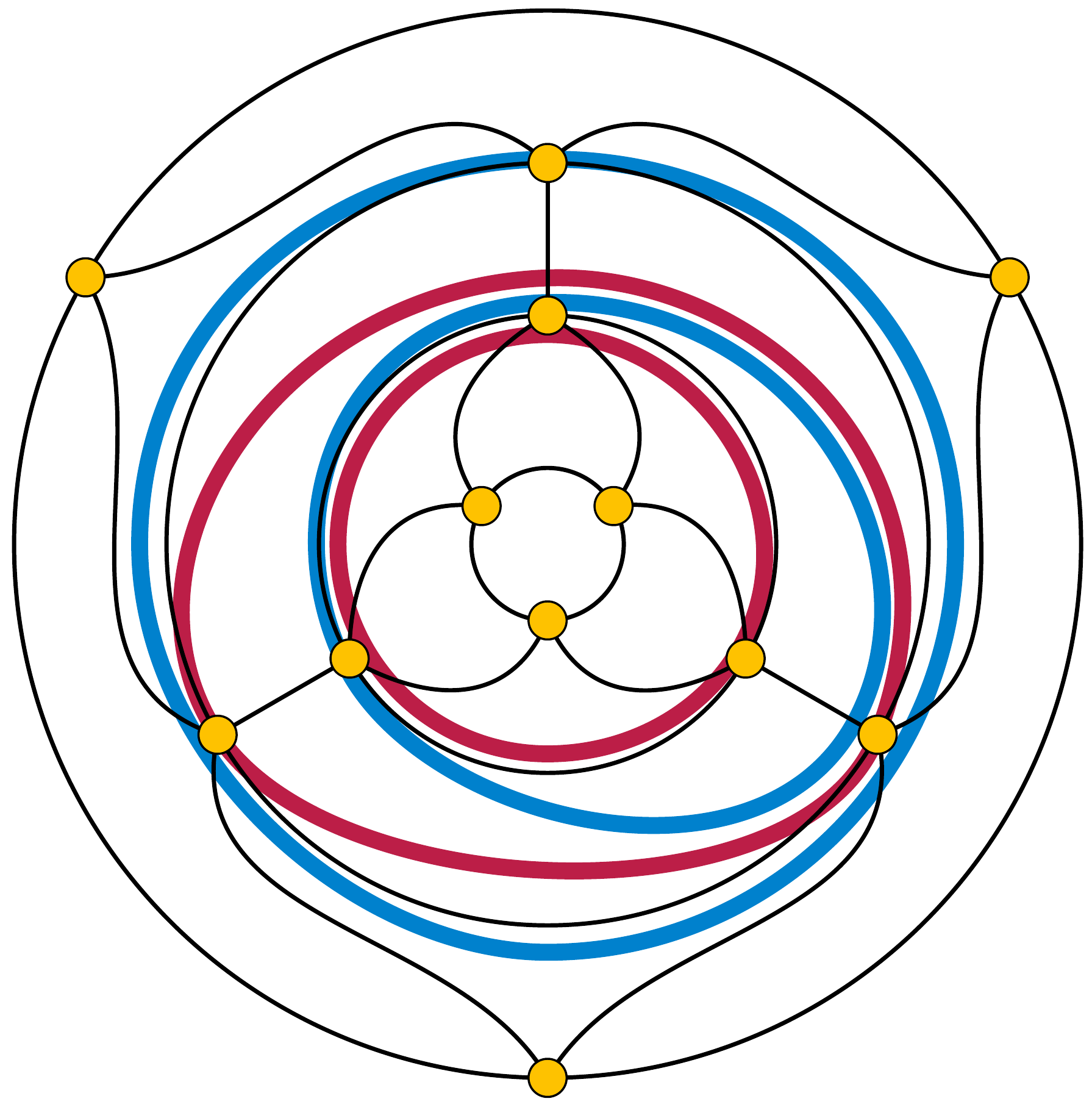}
\caption{A $3$-connected planar graph (yellow vertices and black edges) and a maximal laminar set of four $3$-cutsets (the triples of vertices that the four red and blue curves pass through). There are five other ways of choosing a maximal laminar set of $3$-cutsets in the same graph.}
\label{fig:laminar}
\end{figure}

\subsection{New reults}
In this paper, we describe a linear time  algorithm  to find a maximal set of laminar $3$-cutsets  in a $3$-connected planar graph (\autoref{fig:laminar}).
Our algorithm can also be used to find a maximal set of laminar constrained $3$-cutsets, when we limit the cutsets to belong to one of several families including the following: (i) all 3-cuts, (ii) all non-trivial 3-cuts, and 
(iii) all $v$-non-shiftable $3$-cuts for some vertex~$v$ in such a graph. 
Here, we say that a cutset  $X$ disjoint from $v$ is $v$-non-shiftable  for some vertex~$v$ if it is disjoint from $v$ and every vertex of $X$ has at least two neighbours that are not in the component of $G-X$ containing~$v$. (That is, these neighbours must either be in a different component or in $X$ itself.)

One of the main motivations of our work is that it forms an important subroutine in a recent linear-time algorithm of Kawarabayashi et al.{} for finding disjoint paths in graphs that are not necessarily planar~\cite{KawLiRee-15}. The case of $v$-non-shiftable $3$-cuts is the one needed for this application.

\subsection{Overview}
The basic idea of our algorithm is to follow the following outline:
\begin{itemize}
\item Form a derived planar graph (the \emph{barycentric subdivision}, or incidence graph of vertices, edges, and faces in the unique planar embedding of the given graph) such that $3$-cutsets in the original graph correspond uniquely to certain $6$-cycles in the derived graph, which we call \emph{canonical cycles}.
\item Apply a planar subgraph isomorphism algorithm~\cite{Epp-JGAA-99} to find all the canonical cycles, and all the non-laminar pairs of canonical-cycles, in the derived graph. Construct from them a \emph{conflict graph} in which the vertices are $6$-cycles and the edges are non-laminar pairs.
\item Find and return a maximal independent set of vertices in the conflict graph.
\end{itemize}
\begin{figure}[t]
\centering\includegraphics[scale=0.4]{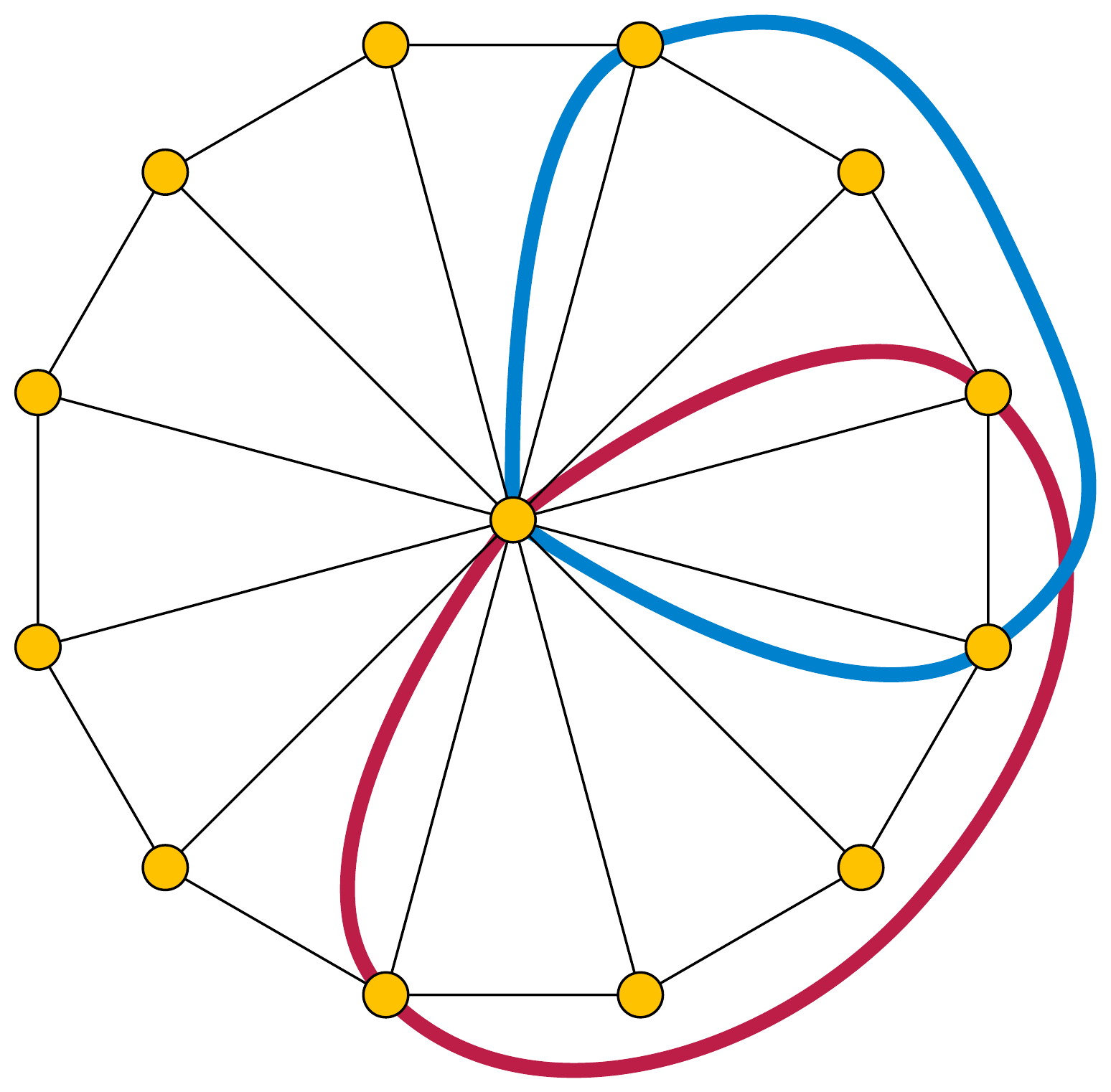}
\caption{A wheel graph has quadratically many $3$-cutsets, formed by the hub and any two non-adjacent vertices of the outer cycle, and quartically many non-laminar pairs of $3$-cutsets.}
\label{fig:wheel}
\end{figure}
However, in its basic form, this algorithm does not take linear time. The issue is that some $3$-connected planar graphs such as the wheels (\autoref{fig:wheel})  may have as many as $\Theta(n^2)$ $3$-cutsets and as many as $\Theta(n^4)$ non-laminar pairs of $3$-cutsets, making the conflict graph too large.
To get around this problem, we show that all of the non-linear complexity of the conflict graph can be attributed to certain disjoint wheel-like structures in the derived graph that we call \emph{frames}.
Using these structures gives us a refined algorithm outline that performs the following steps.
\begin{itemize}
\item Construct the derived graph as before.
\item Apply a planar subgraph isomorphism algorithm to find all maximal frames in the derived graph.
\item Separately, for each frame, find a nearly-maximal laminar family of  canonical cycles that cannot conflict with the cycles for any other frame.
\item Cut the derived graph along all of these $6$-cycles into subproblems within which there are no frames, and within which all $6$-cycles are laminar with the ones already chosen for the frames.
\item Construct the conflict graph for the union of the subproblems and find a maximum independent set in this conflict graph.
\item Return the union of the $3$-cutsets found within each subproblem and the $3$-cutsets found within each frame.
\end{itemize}
In the following sections we expand on these ideas.

\section{The basic algorithm}

In our description of our algorithm we will occasionally refer to the inside or the outside of a cycle in our embedded graph.  In order for this language to make sense, we choose (arbitrarily for now) one face of a planar graph to be the ``outer'' face. With this choice, the outside of a cycle in the graph is the side that contains the outer face, and the inside is the side that does not contain the outer face. The vertices and edges of the cycle itself are neither inside nor outside. One cycle contains another one laminar to it if the contained cycle includes edges from the inside of the containing cycle.

\subsection{From cutsets to cycles}
\label{sec:construct-dg}

We begin with some basic properties of $3$-cutsets in $3$-connected planar graphs.

\begin{definition}
A $3$-cutset in a $3$-connected graph is a subset of three vertices whose deletion would disconnect the remaining graph.
A cutset is \emph{trivial} if it consists of the three neighbours of a degree-three vertex, and \emph{non-trivial} otherwise.
\end{definition}

\begin{figure}[t]
\centering\includegraphics[scale=0.4]{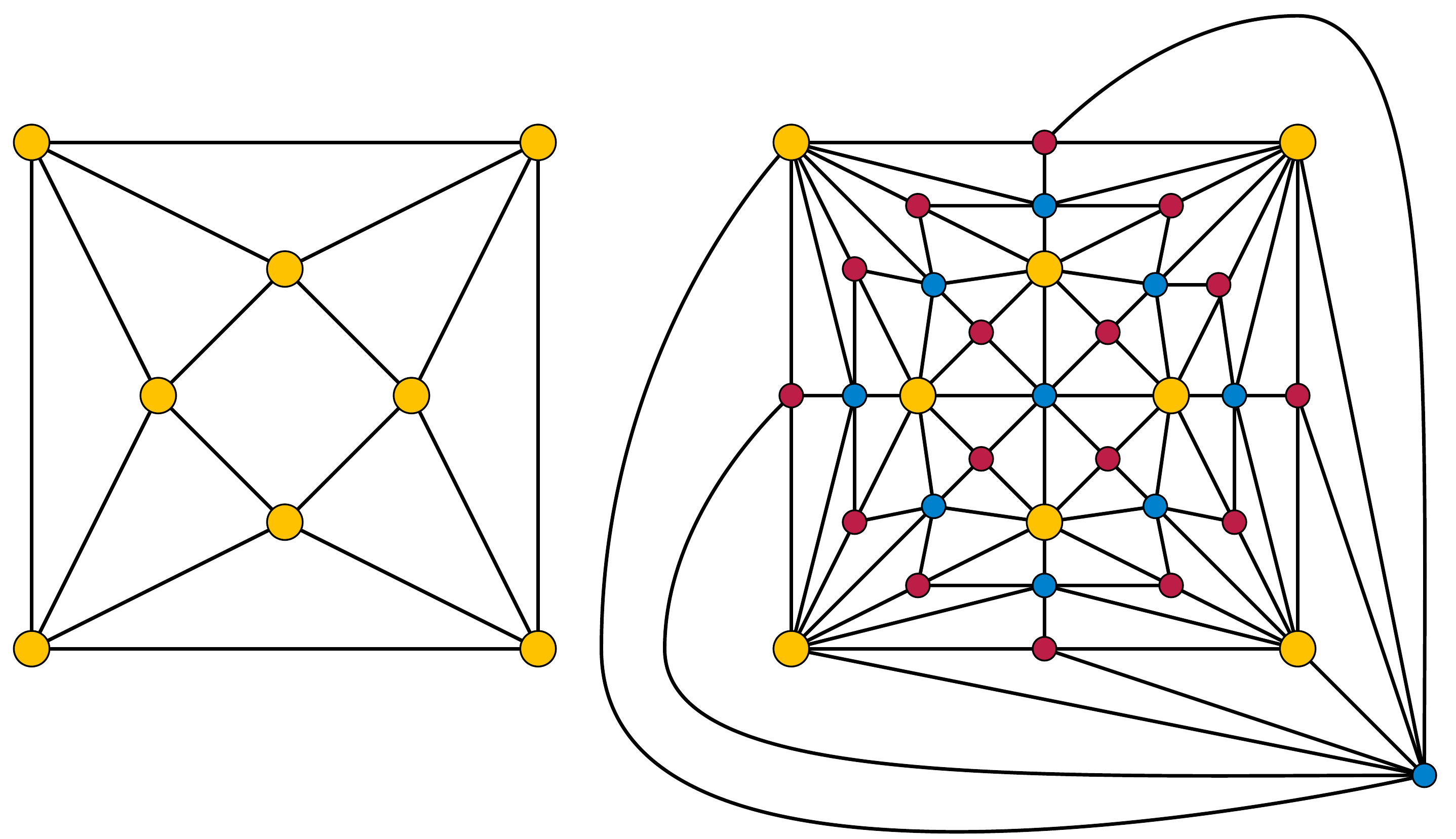}
\caption{A graph $G$ and its derived graph $\derivedgraph(G)$}
\label{fig:derivedgraph}
\end{figure}

Rather than seeking $3$-vertex cuts in the given $3$-connected planar graph $G$ directly, we will instead seek certain constrained $6$-cycles in the \emph{barycentric subdivision} $\derivedgraph(G)$ (\autoref{fig:derivedgraph}). This is defined as follows.

\begin{definition}
The \emph{barycentric subdivision} $\derivedgraph(G)$ of a $3$-connected planar graph $G$ has a vertex for each vertex, edge, or face of the unique planar embedding of $G$. It has an edge for each incident vertex-edge, vertex-face, or edge-face pair in~$G$.
\end{definition}

The resulting graph $\derivedgraph(G)$ is planar, with a planar embedding inherited from~$G$.  Each face of $\derivedgraph(G)$ is a triangle, and corresponds to a \emph{flag} in $G$, an incident triple of a vertex, edge, and face. Clearly, it can be constructed from $G$ in linear time.

\begin{definition}
A \emph{canonical cycle} is an induced $6$-cycle in  $\derivedgraph(G)$ that alternates between vertices of $G$ and non-vertices of $G$ such that, whenever two vertices $u$ and $v$ of $G$ belong to the cycle and are adjacent in~$G$, the canonical cycle follows the path $u$--$uv$--$v$ in $\derivedgraph(G)$. Additionally, we require that the six vertices of the induced cycle not be the six neighbours of a triangular face of $G$.
\end{definition}

\begin{figure}[t]
\centering\includegraphics[scale=0.4]{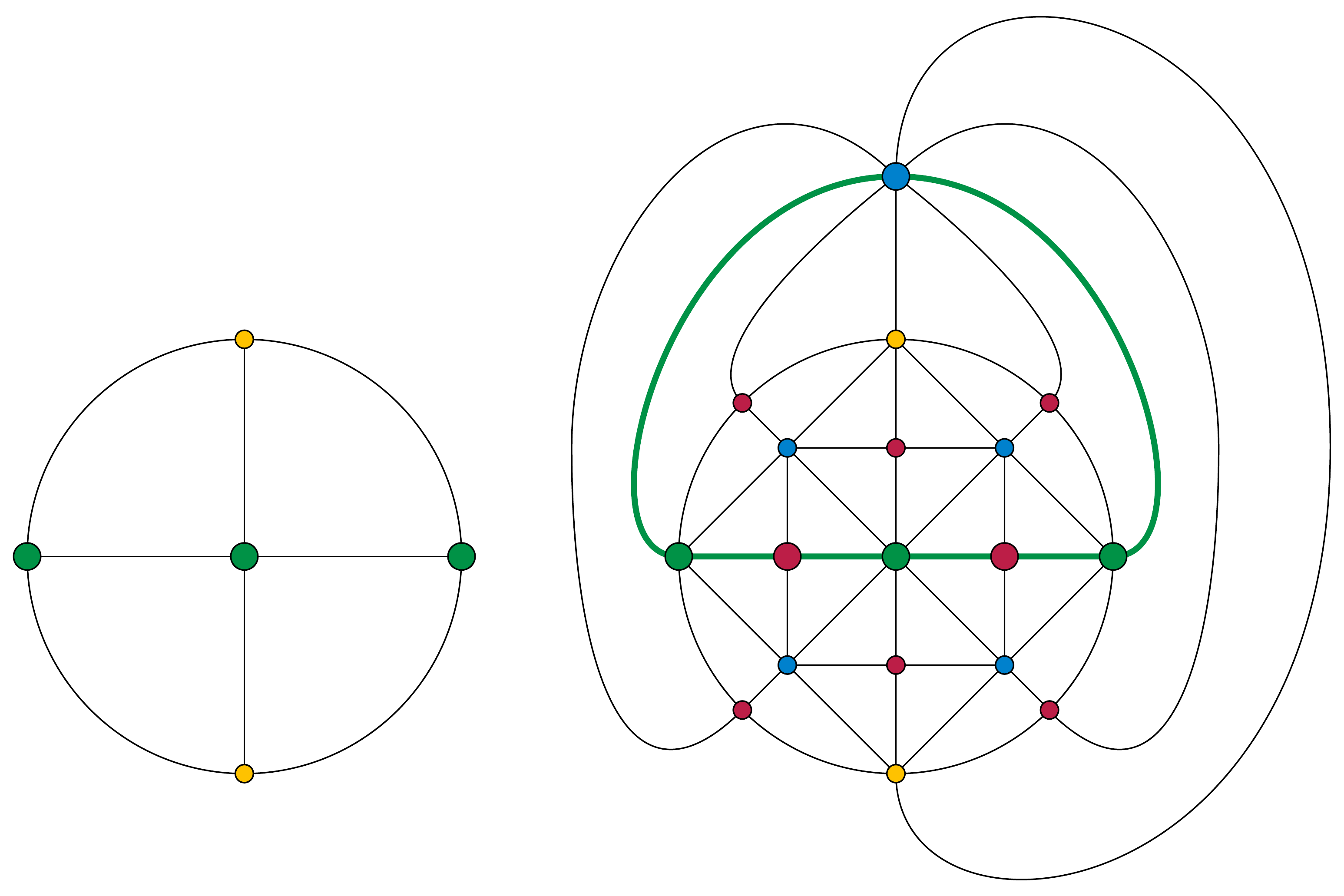}
\caption{A separating triple in a graph $G$ (the three larger green vertices) and the corresponding canonical cycle in the derived graph $\derivedgraph(G)$ (the heavy green cycle)}
\label{fig:canonical-cycle}
\end{figure}

An example of a 3-separation and the corresponding canonical cycle is depicted in \autoref{fig:canonical-cycle}.

\begin{lemma}
\label{lem:cutset-cc-bijection}
The $3$-cutsets of a given $3$-connected planar graph $G$ correspond one-for-one with the canonical cycles of $G$. Each $3$-cutset of $G$ forms three of the six vertices of a canonical cycle, and each canonical cycle is derived from a $3$-cutset in this way.
\end{lemma}

\begin{proof}
A triple of vertices can all belong to at most one canonical cycle, for if any two of the vertices are connected by an edge in $G$ then the cycle must pass through the vertex of $\derivedgraph(G)$ corresponding to that edge, and if there is no such edge then there can be at most one face of $G$ containing both vertices by the assumption of $3$-connectivity. So for each pair of vertices in the triple, there is only one two-edge path in $\derivedgraph(G)$ that connects that pair and can be part of a canonical cycle.

Every $3$-cutset corresponds to a canonical cycle. For, if $X$ is a $3$-cutset, $G-X$ must have exactly two connected components: each component attaches to all three vertices of $X$, else $G$ would not be $3$-connected, and more than two components could be contracted to form a $K_{3,3}$ minor, contradicting the planarity of $G$. Contracting each component of $G-X$ to a single vertex produces a drawing of $K_{3,2}$ (possibly with some additional edges connecting pairs of vertices in $X$) in which one can draw a simple closed curve that alternates between vertices of $X$ and faces of the drawing and that separates the two contracted components. The contraction process can then be reversed to produce a curve through the drawing of $G$ that again alternates between vertices of $X$ and faces of the drawing and that separates the two components of $G-X$. If this is not already a canonical cycle, it can be made into one by replacing some of the faces by edges between pairs of vertices in $X$.

The three vertices in $G$ of every canonical cycle cannot all be incident to one face $f$ of $G$, by the following case analysis:
\begin{itemize}
\item A $6$-cycle through the three vertices and $f$ would not be induced.
\item A $6$-cycle through the three vertices and any other face $f'$ would not be canonical, because it should instead use the edge separating $f$ from $f'$.
\item A $6$-cycle through the three vertices and no faces would form a triangle surrounding $f$, 
and therefore would not be canonical.
\end{itemize}
Therefore, both inside and outside the canonical cycle there must exist a path with at least one interior vertex separating the face containing two of the canonical cycle vertices from the third vertex. The three vertices then separate the interior vertices of the paths inside the canonical cycle from those outside the canonical cycle, so every canonical cycle corresponds to a $3$-cutset.
\end{proof}

\subsection{Laminarity}

We formalize the concept of laminar and non-laminar pairs of $3$-cutsets and canonical cycles in the following definitions.

\begin{definition}
Two abstract sets $S$ and $T$, subsets of a universe $U$, are laminar if $S\subset T$, $T\subset S$, or $S\cap T=\emptyset$, and non-laminar if all three of the sets $S\setminus T$, $T\setminus S$,  and $S\cap T$ are nonempty.
Two $3$-cutsets $X$ and $Y$ of a $3$-connected planar graph $G$ are non-laminar if $X$ separates two vertices of $Y$ (that is, $Y$ includes vertices in both of the components of $G-X$) or if $Y$ separates two vertices of $X$; otherwise, they are laminar.
Two canonical cycles $X$ and $Y$ of $\derivedgraph(G)$ are laminar if the sets of faces of $\derivedgraph(G)$ interior to $X$ and to $Y$ (viewed as subsets of the set of all faces of $\derivedgraph(G)$) are laminar, and non-laminar if these two sets of faces are non-laminar.
\end{definition}

By the following observation, we can find laminar cutsets by searching for laminar canonical cycles.

\begin{observation}
Any two $3$-cutsets of a $3$-connected planar graph $G$ are laminar if and only if the corresponding two canonical cycles of $\derivedgraph(G)$ are laminar.
\end{observation}

\subsection{Finding small subgraphs}

Both in the basic version of our algorithm (in which we construct the conflict graph of canonical $6$-cycles) and in the full version (which also involves finding wheel-like subgraphs and using them to partition the problem into simpler subproblems) we need to solve a form of the planar subgraph isomorphism problem. In this problem, we are given a small \emph{pattern} graph $H$ and a larger \emph{text} graph $G$, and must list all subgraphs of $G$ that are isomorphic to $H$.
We apply a known linear-time solution for this problem:

\begin{lemma}[Eppstein~\cite{Epp-JGAA-99}]
\label{lem:sgi}
Let $G$ and $X$ be planar graphs, both of whose vertices have been assigned (not necessarily distinct) labels.
Let $n=|V(G)|$ (considered to be variable), and $x=|V(X)|$.
Then there is a computable function $f$ such that it is possible to list all instances of $X$ as a labeled subgraph of $G$, or as an induced labeled subgraph of $G$, in time $O((n+k)f(x))$, where $k$ is the number of instances to be listed. The method extends to partially induced subgraphs in which some pairs of nonadjacent vertices in $X$ must correspond to nonadjacent vertices in~$G$, and other pairs are not so constrained.
\end{lemma}

The solution technique involves using groups of consecutive layers of a breadth-first search tree to cover $G$ by a collection of subgraphs of bounded treewidth, with total linear size, such that every subgraph of $G$ with diameter at most $x$ is entirely contained in one of these bounded-treewidth subgraphs. Then, standard dynamic programming methods within each bounded-treewidth subgraph may be used to find all copies of $X$ that belong to that subgraph and do not belong to earlier subgraphs in the collection.
Since its original publication, this algorithm has been improved to have only single-exponential dependence on the parameter~$x$~\cite{Dor-STACS-10}, and it has been greatly generalized to the problem of testing any first-order property in classes of graphs with bounded expansion~\cite{NesOss-S-12}.

The same methods can be used in the same way to prove the following variation, in which the subgraphs that we list as output are part of larger patterns:

\begin{lemma}
\label{lem:sgi2}
Let $G$ be a labeled planar graph with $n$ vertices, $X$ be a labeled planar graph with $x$ vertices, let $Y$ be a subgraph of $X$, and consider $n$ to be variable and $x$ to be constant. Then it is possible to list all placements of $Y$ as a subgraph of $G$ that can be extended to placements of $X$ as a subgraph of $G$, in time proportional to $n$ and to the number of placements of $Y$ produced as output.
\end{lemma}

This allows us to find canonical cycles efficiently:

\begin{lemma}
\label{lem:list-canonical-cycles}
It is possible to list all canonical cycles in $\derivedgraph(G)$ in time proportional to its size and number of canonical cycles.
\end{lemma}

\begin{figure}[t]
\centering\includegraphics[scale=0.4]{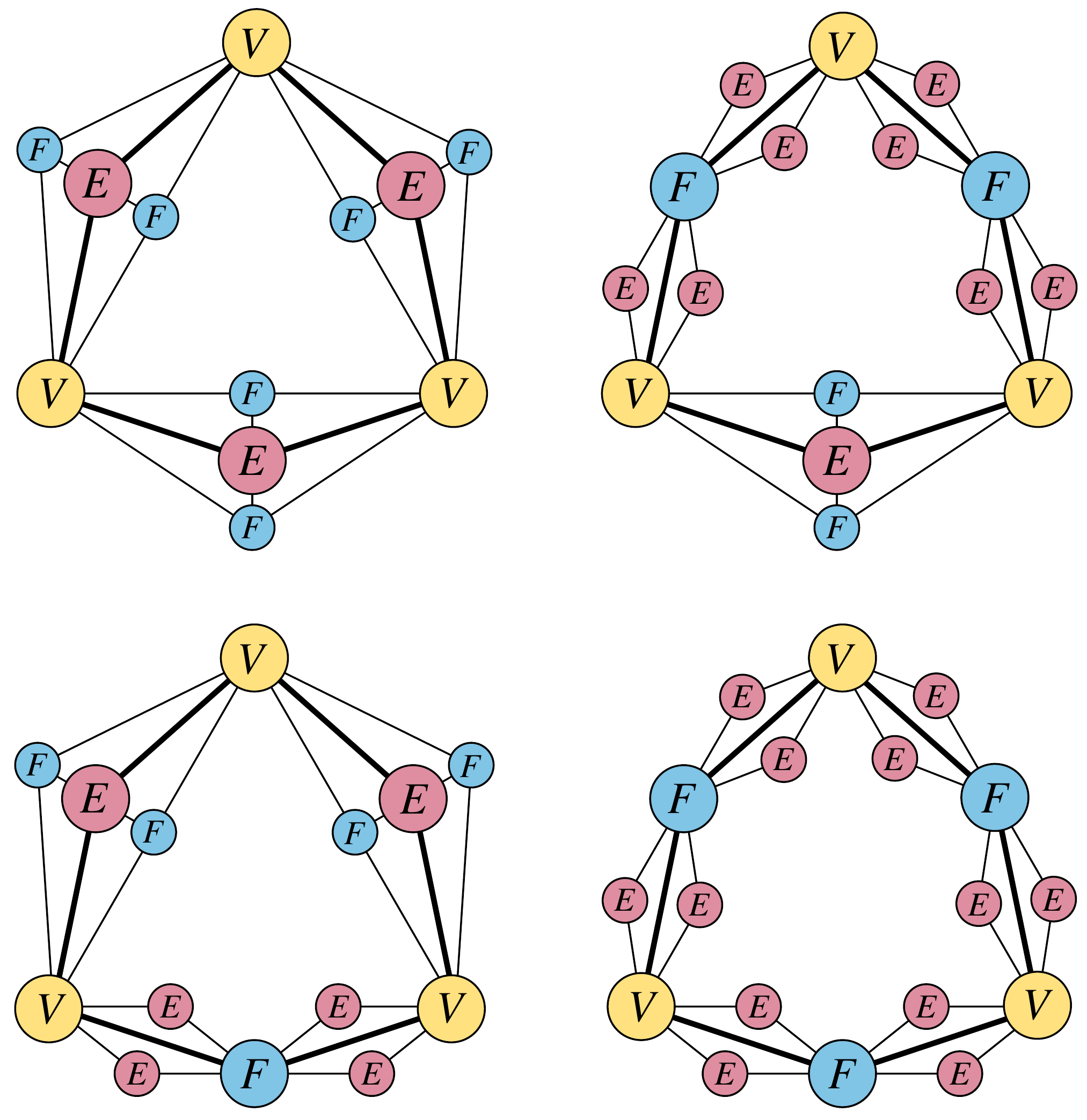}
\caption{The four possible expansions of a $(V,E,F)$-labeled canonical cycle $X$ (larger vertices and thicker edges) into an augmented subgraph $Y$ of $\derivedgraph(G)$, for the proof of \autoref{lem:list-canonical-cycles}.}
\label{fig:canonical-patterns}
\end{figure}

\begin{proof}
We will show that this can be solved using \autoref{lem:sgi2}.
Each canonical cycle in $\derivedgraph$ will be the labeled planar graph $X$ of the lemma, with the three labels $V$, $E$, and $F$ being used to distinguish vertices of $\derivedgraph(G)$ that correspond to vertices, edges, and faces of $G$ respectively. A canonical cycle is a $6$-cycle whose labels alternate between $V$ and non-$V$ vertices; there are four patterns of labels according to how many $E$ and $F$ labels there are in the cycle. We will treat each of these four patterns as the subgraph $X$ in four instances of \autoref{lem:sgi2}, augmented to be part of a larger subgraph $Y$ whose added vertices and edges prevent us from listing $6$-cycles that are not canonical. If $X$ is any labeled $6$-cycle that alternates between vertices labeled $V$ and non-$V$, we augment $X$ to form $Y$ as follows:
\begin{itemize}
\item For each vertex of $X$ labeled $E$, we add two vertices labeled $F$, adjacent both to the $E$ vertex and its two neighbours in the cycle.
\item For each vertex of $X$ labeled $F$, we add four vertices labeled $E$, two of which are adjacent to it and its clockwise neighbour in the cycle, and the other two of which are adjacent to it and its counterclockwise neighbour.
\end{itemize}
The four possible expansions of a canonical cycle, formed in this way, are depicted in \autoref{fig:canonical-patterns}.
In any canonical cycle of $\derivedgraph(G)$, these vertices will necessarily exist (as the neighbouring edges or faces in $G$ to the faces or edges of the canonical cycle) and be distinct from each other: any coincidence between two added $F$-vertices would mean that the $6$-cycle is not induced or the boundary cycle of a triangular face, and any coincidence between two added $E$-vertices would mean that the $6$-cycle is not induced or goes through a face when it could go through an edge, none of which is allowed in a canonical cycle. Conversely, any $6$-cycle in $\derivedgraph(G)$ that can be extended to a subgraph matching one of these augmented $Y$ graphs is necessarily canonical. Therefore, all canonical cycles can be found by combining four instances of \autoref{lem:sgi2}, one for each $(X,Y)$ pair.
\end{proof}

To find non-trivial or $v$-non-shiftable cutsets, we will use the same method to find all canonical cycles, and then filter out the trivial or shiftable ones as an extra postprocessing step. (It would also be possible to describe larger patterns that could be used to find these cutsets directly, without the filtering, but this would require a more complicated case analysis and would not speed up our overall algorithm.)

In order to find non-laminar pairs of cycles, we need a variation of the same subgraph isomorphism technique that can find subgraphs whose embedding (as induced from the unique embedding of the given $3$-connected planar graph) is of a given type. This is not a first-order graph property, but it can also be tested in the same time bounds, as an instance of the \emph{planar subdrawing equivalence} problem defined by Dorn~\cite{Dor-STACS-10}. We state this result as the following lemma.

\begin{lemma}[Dorn~\cite{Dor-STACS-10}]
\label{lem:sgi3}
Let $G$ be a labeled plane graph with $n$ vertices, $X$ be a labeled plane graph with $x$ vertices, and consider $n$ to be variable and $x$ to be constant. Then it is possible to list all instances of $X$ as a labeled subgraph of $G$, with the induced embedding of the subgraph matching the given embedding of $X$, in an amount of time that is linear in~$n$ and in the number of copies of $X$. 
\end{lemma}

To use this to find non-laminar pairs, we consider all of the ways in which two of the subgraphs from \autoref{fig:canonical-patterns} can overlay each other to form a single embedded non-laminar pair, and apply \autoref{lem:sgi3} to find all instances in $\derivedgraph(G)$ of each of these patterns. We omit the (finite but messy) case analysis.

\subsection{Constructing the conflict graph}

Given a vertex-labeled graph $H$ (either the derived graph $\derivedgraph(G)$ or one of its subgraphs) we define the \emph{conflict graph} $\conflictgraph(H)$ as follows.

\begin{definition}
The \emph{conflict graph} $\conflictgraph(H)$ of a graph $H$ is another graph, with vertices in one-to-one correspondence with the canonical cycles in $H$. Two vertices in $\conflictgraph(H)$ are adjacent if and only if they correspond to a non-laminar pair of canonical cycles.
\end{definition}

We can use planar subgraph isomorphism techniques to construct $\conflictgraph(H)$.

\begin{lemma}
\label{lem:construct-IG}
We can construct $\conflictgraph(H)$ in time linear in the sum of the sizes of the input and output graphs.
\end{lemma}

\begin{proof}
We use \autoref{lem:list-canonical-cycles} to find all canonical cycles of $H$, and \autoref{lem:sgi3} to find all embedded subgraphs in the form of two non-laminar canonical cycles. This produces an object for each vertex and edge of $\conflictgraph(H)$. An adjacency list representation of $\conflictgraph(H)$, matching each vertex to its incident edges and vice versa, can be constructed by labeling the vertices of $H$ by distinct integers from $1$ to~$n$ (where $n$ is the total number of vertices), labeling each canonical cycle by a cyclically reduced $6$-tuple of these vertex labels,  using radix sort to jointly sort the vertices and edges of $\conflictgraph(H)$ (with two copies of each edge) by the labels of their canonical cycles, and finding incident pairs of vertices and edges of $\conflictgraph(H)$ adjacent to each other in the sorted order.
\end{proof}

This gives us the main algorithm that we will apply to find maximal laminar sets of cycles, after decomposing the problem into subproblems within which $\conflictgraph(H)$ can be proved to be linear.

\begin{lemma}
\label{lem:solve-frameless}
Given a vertex-labeled graph $H$, we can find a maximal laminar set of canonical cycles in $H$ in time linear in the sum of the sizes of $H$ and of $\conflictgraph(H)$.
The same technique applies to any subset of the canonical cycles that can be identified by the subgraph isomorphism techniques of \autoref{lem:sgi2} or \autoref{lem:sgi3}, and in particular to the canonical cycles corresponding to nontrivial or $v$-non-shiftable $3$-cutsets.
\end{lemma}

\begin{proof}
We use \autoref{lem:construct-IG} to construct $\conflictgraph(H)$, and then apply a linear-time greedy algorithm that constructs a maximal independent set in $\conflictgraph(H)$ by considering each vertex one by one (in an arbitrary order), including it in the independent set if and only if none of its neighbours has been already included. The times to construct $\conflictgraph(H)$ and to perform the greedy maximal independent set algorithm are both linear in the sum of the sizes of the graphs. The canonical cycles in $H$ that correspond to the vertices of the maximal independent set form a maximal laminar set of cycles. In the cases of nontrivial or $v$-non-shiftable $3$-cutsets, we find a maximal independent set in an induced subgraph of $\conflictgraph(H)$ that includes only the relevant types of cutsets, rather than the whole conflict graph.
\end{proof}

\section{Sparsifying the conflict graph}

In this section we describe how to decompose $\derivedgraph(G)$ into subgraphs of two types: \emph{frames}, wheel-like substructures containing many non-laminar pairs of canonical cycles, and subgraphs without frames, in which the conflict graph can be shown to be sparse. The decomposition will have the properties that all canonical cycles of $\derivedgraph$ can be found in one of these subgraphs, and that no two canonical cycles from different subgraphs can be non-laminar. Therefore, after decomposing the graph in this way, we will be able to find a maximal set of canonical cycles separately in each subgraph.

\subsection{Allowable cycles}

In performing this decomposition, it is convenient to consider, instead of the canonical cycles of $\derivedgraph(G)$, a different type of cycle, which we call \emph{allowable cycles}. The allowable cycles will still correspond to $3$-cutsets of $G$, but the correspondence is not one-to-one: each non-trivial $3$-cutset can be represented uniquely as an allowable cycle, but each allowable cycle can represent $O(1)$ different $3$-cutsets. In exchange for this ambiguity, the ways that allowable cycles can cross each other will be more restricted, making our case analysis simpler.

\begin{definition}
In~$\derivedgraph(G)$, we define an \emph{allowable cycle} to be an induced $6$-cycle that avoids the following forbidden configurations:
\begin{enumerate}
\item paths of the form $x$--$f$--$y$, where $x$ and $y$ are adjacent vertices of $G$ that both belong to the same face $f$ of $G$, and
\item paths of the form $f_1$--$x$--$f_2$ (where $f_1$ and $f_2$ are the two faces of $G$ that share edge $xy$).
\item cycles in which all six vertices have a common neighbour.
\end{enumerate}
\end{definition}

\begin{figure}[t]
\centering\includegraphics[scale=0.5]{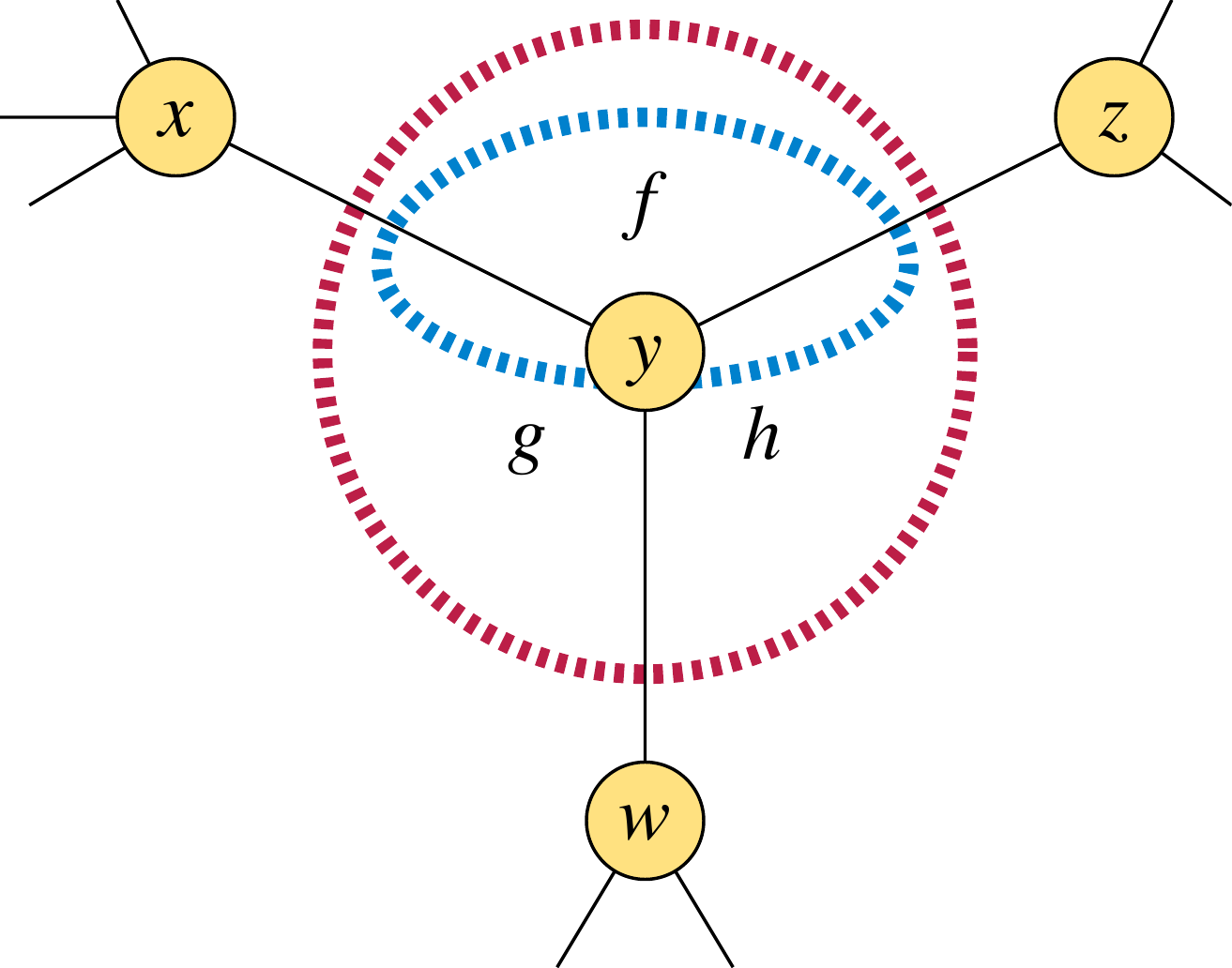}
\caption{Illustration for \autoref{lem:path-through-adjacent-edges}. An allowable cycle cannot pass through the two consecutive edges $xy$ and $yz$ of face $f$, for to do so it would have to either be completed through $y$ (inner dashed blue curve), making it a non-induced cycle, or through an edge $yw$ (outer dashed red curve), making its six vertices have the common neighbour~$y$.}
\label{fig:allowable-nonconsecutive}
\end{figure}

Another forbidden configuration is not part of the definition, but follows automatically from it:

\begin{lemma}
\label{lem:path-through-adjacent-edges}
If $f$ is a face of $G$, and $C$ is an allowable cycle containing $f$, then the two neighbours of $f$ in $C$ cannot be edges that share a common vertex.
\end{lemma}

\begin{proof}
Suppose that $xy$ and $yz$ are edges of $f$, and that the faces on the other sides of these edges from~$f$ are $g$ and $h$ respectively; these features of $G$ are illustrated in \autoref{fig:allowable-nonconsecutive}.  In any allowable cycle passing through $xy$--$f$--$yz$, the cycle must continue on both sides as $g$--$xy$--$f$--$yz$--$h$, for continuing instead through the only other neighbours of $xy$ and $yz$ in $\derivedgraph(G)$ (the vertices $x$, $y$, and $z$) would cause the cycle to be non-induced, as these vertices are all incident to $f$. The sixth vertex of the cycle must be a feature of $G$ that is a common neighbour of faces $g$ and $h$, and this can only be either the vertex $y$ or another edge $yw$. But neither of these choices completes an allowable cycle: using~$y$ would create a cycle that is not induced (because $y$ is incident to $f$; the inner blue cycle of the figure) while using $yw$ would create a cycle in which all six vertices have the common neighbour~$y$ (the outer red cycle of the figure).
\end{proof}

We now describe a general method of converting allowable cycles into canonical cycles.

\begin{definition}
If $C$ is an allowable cycle, we define a \emph{shift} of $C$ to be the cycle formed by replacing each path in $C$ of the form $f_1$--$e$--$f_2$ (where $e$ is an edge of $G$ and $f_1$ and $f_2$ are the two incident faces) by one of the two paths of the form $f_1$--$x$--$f_2$ or $f_1$--$y$--$f_2$ (where $x$ and $y$ are the two endpoints of $e$), and then replacing each path in $C$ of the form $u$--$f$--$v$ where $u$ and $v$ are adjacent in $G$ by the path $u$--$uv$--$v$.
\end{definition}

\begin{observation}
Every allowable cycle has at most eight shifts: each shift is defined by the choice of which endpoint to use to replace one or more edges in the cycle, there are two choices per replaced edge, and the replaced edges are non-adjacent in the cycle so there are at most three replaced edges.
\end{observation}

Shifts do not always produce canonical cycles, but the exceptions are highly constrained:

\begin{lemma}
\label{lem:shift-canonical}
Every shift $C'$ of an allowable cycle $C$ is either a canonical cycle or the set of six neighbours of a triangular face in $G$.
\end{lemma}

\begin{proof}
Because allowable cycles are induced cycles, every edge $e$ of $G$ that belongs to $C$ has as its two neighbours in $C$ either the two endpoints of $e$ or the two faces separated by $C$; in the latter case, $e$ is replaced by a vertex of $G$ in $C'$. Therefore, for every two consecutive vertices of $C'$, exactly one of the two corresponds to a vertex of $G$, either because the two corresponding consecutive vertices of~$C$ already included a vertex of $G$ or because they were an incident edge-face pair in $G$ of which the edge was replaced in $C'$ by a vertex. Therefore, $C'$ alternates between vertices of $G$ and non-vertices, and the three vertices are all distinct by \autoref{lem:path-through-adjacent-edges}.

\begin{figure}[t]
\centering\includegraphics[scale=0.5]{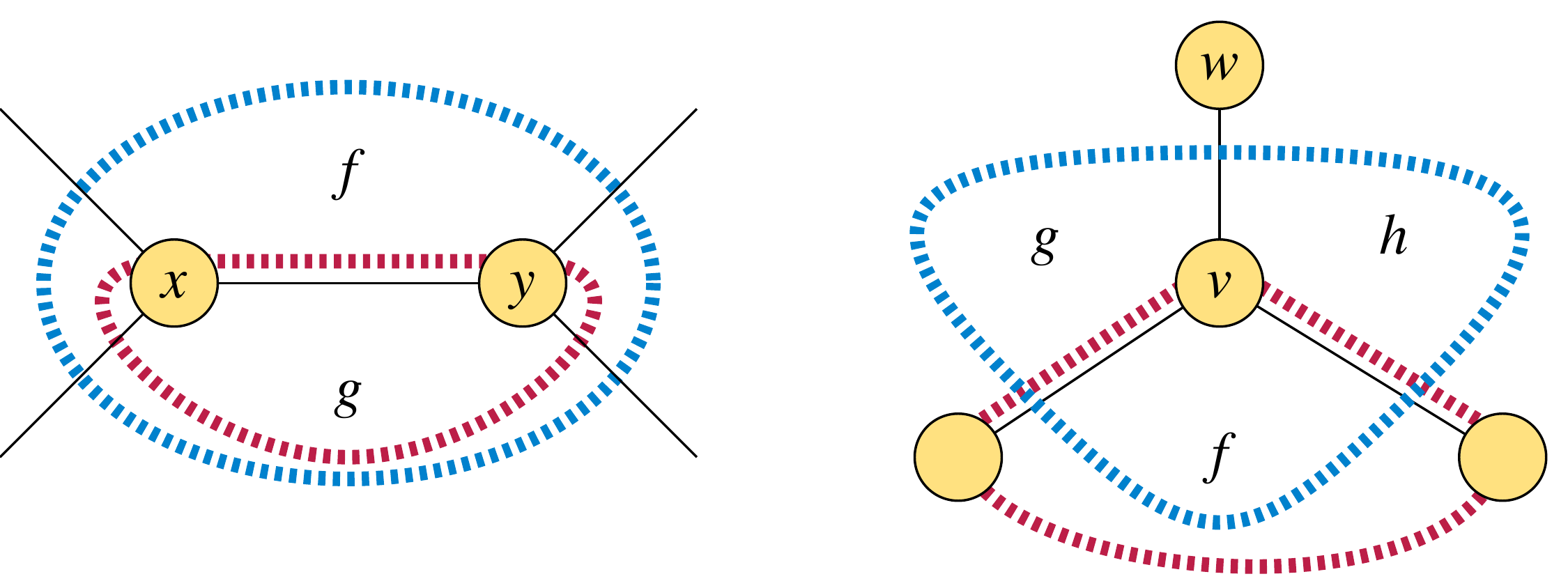}
\caption{Cases for \autoref{lem:shift-canonical}, showing a hypothetical allowable cycle $C$ (blue) and its shift (red). Left: $C'$ cannot contain an incident pair of an edge $xy$ and face $g$, where $xy$ is shifted from face $f$, because then only one of the two endpoints $x$ and $y$ could be included in $C'$ as the shift of edge $xy$ in $C$. Right: $C'$ cannot contain an incident but non-consecutive pair of a vertex $v$ and face $f$, because $v$ could only be included as the shift of an edge $vw$ (else $C$ would be non-induced) and then $C$ would necessarily pass through the edges separating $f$ from the faces $g$ and $h$ on either side of $vw$, violating \autoref{lem:path-through-adjacent-edges}.}
\label{fig:shift-canonical}
\end{figure}

Whenever two vertices $x$ and $y$ of $G$ in $C'$ are adjacent, the vertex between them in $C'$ must correspond to the edge $xy$ in $G$, because paths from $x$ to $y$ through a face are replaced by paths through this edge as part of the definition of the shift.

$C'$ cannot include any incident pair of an edge and a face in $G$. For, if both elements of the pair also belonged to $C$, they would have caused it to be non-induced, violating the definition of an allowable cycle. If, on the other hand, $C'$ includes an edge $xy$ that was shifted from a face $f$, and also includes a face $g$ incident to $xy$ (\autoref{fig:shift-canonical}, left), then (as $f$ and $g$ share the edge $xy$) to be an allowable cycle, $C$ would have to include the sequence $f$--$xy$--$g$. In this case one of $x$ or $y$ (without loss of generality $x$) would be included in $C'$ as the shift of the edge $xy$, but the other vertex $y$ could not be included: it could not have been part of $C$ because then the incidence $y$--$xy$ would cause $C$ to be non-induced, and it could not have been added to $C'$ as the shift of another edge, different from $xy$, by \autoref{lem:path-through-adjacent-edges}.

$C'$ cannot include a non-consecutive but incident pair of a vertex and edge in $G$, because for each vertex of $G$ in $C'$, the only non-consecutive non-vertex feature of $G$ in $C'$ is the feature connecting the other two vertices, which we have already argued are distinct. It also cannot include a non-consecutive but incident pair of a vertex $v$ and face $f$ in $G$, because then all three vertices of $C'$ would be incident to $f$. If $v$ were part of $C$, then $C$ would be non-induced.
Otherwise, if~$v$ was shifted from an edge $vw$ in $C$, and $vw$ separates faces $g$ and $h$, then $C$ must pass through the sequence $g$--$vw$--$h$ (\autoref{fig:shift-canonical}, right). The other two vertices of $G$ in $C'$ must be on edges that separate $g$ and $h$ from $f$, and to be an allowable cycle $C$ must connect $f$ to $g$ and to $h$ through these edges; but this violates \autoref{lem:path-through-adjacent-edges}. Thus, since none of the possible cases for a non-consecutive but incident pair can occur in $C'$, it must be the case that $C'$ forms an induced cycle.

Thus, all the conditions from the definition of a canonical cycle, except the condition that~$C'$ cannot be the set of six neighbours of a triangular face of $G$, are met. 
\end{proof}

We saw earlier that canonical cycles correspond one-for-one with separating triples. In particular, each separating triple can be represented by a canonical cycle. We now prove a corresponding result for allowable cycles.

\begin{lemma}
\label{lem:canonical-allowable}
Let $xyz$ be a separating triple of $G$. Then there is an allowable cycle $C$ in $\derivedgraph(G)$ such that $x$, $y$, and $z$ are vertices of a shift of~$C$.
\end{lemma}

\begin{proof}
Draw a Jordan curve in the plane that passes through $x$, $y$, and $z$, is otherwise disjoint from~$G$, and separates the two components of $G\setminus\{x,y,z\}$; let $f$, $g$, and $h$ be the three faces crossed by this curve. Form a $6$-cycle $C$ in $\derivedgraph(G)$ by starting with the six objects in cyclic order $x,f,y,g,z,h$, replacing $f$, $g$, or $h$ by the edge $xy$, $yz$, or $xz$ if those  edges belong to~$G$, and then replacing $x$, $y$, or $z$ by a shared edge between two of the faces $f$, $g$, or $h$ if both faces remain in the cycle and share an edge. Then, $C$ is necessarily a cycle in $\derivedgraph(G)$. It is an induced cycle, because the existence of any chord of the cycle would imply that $x$, $y$, and $z$ are co-facial and therefore non-separating. And the replacement of vertices or faces of $G$ by edges in $C$, when possible, prevents any of the forbidden paths from occurring. Therefore, $C$ is an allowable cycle whose shift includes $x$, $y$, and $z$.
\end{proof}

Note however that a separating triple may be represented by multiple different allowable cycles.

\begin{lemma}
\label{lem:unique-2-path}
For any two vertices $x$ and $y$ in $\derivedgraph(G)$ there can be at most one way
to connect $x$ to $y$ by a two-edge path that is part of an allowable cycle.
\end{lemma}

\begin{proof}
We consider the possible types of object in $G$ corresponding to vertices $x$ and $y$:
\begin{itemize}
\item If $x$ and $y$ both correspond to vertices of $G$ then these vertices can be connected either by a single face (in which case the vertex of $\derivedgraph(G)$ corresponding to that face is the only possible middle vertex of the path) or by two faces that share an edge (in which case the vertex of $\derivedgraph(G)$ corresponding to that edge is the only possible middle vertex of the path).
\item If $x$ and $y$ both correspond to edges of $G$ then those two edges can be non-incident edges of a common face (in which case the vertex of $\derivedgraph(G)$ corresponding to that face is the only possible middle vertex of the path), they may be non-consecutive edges incident to the same vertex (which gives the only possible middle vertex of the path), or they may share both a vertex and a face. However, in this final case there is no allowable cycle that uses the shared face by \autoref{lem:path-through-adjacent-edges}.
\item If $x$ and $y$ both correspond to faces of $G$ then these faces can share either a single vertex, or an edge and its two endpoints, and in either case there is only one choice for the middle vertex of the path.
\item If $x$ corresponds to a vertex of $G$ and $y$ corresponds to an edge of $G$ then the middle vertex of the path must correspond to a face of $G$. There can only be one face incident to this vertex and disjoint edge without violating the assumption that $G$ is 3-connected.
\item If $x$ corresponds to a vertex of $G$ and $y$ corresponds to a face of $G$ then there can be no allowable cycle that uses a two-edge path from $x$ to $y$, for all two-edge paths must pass through a middle vertex corresponding to an edge of $G$ and would not be induced paths.
\item If $x$ corresponds to an edge of $G$ and $y$ corresponds to a face of $G$ then the middle vertex of the path must correspond to a vertex of $G$, incident to both the edge and the face. There can only be one such vertex, else $x$ and $y$ would be adjacent in $\derivedgraph(G)$ and could not be two steps apart in an induced cycle.
\end{itemize}
\end{proof}

\subsection{Frames}
 
Our analysis focuses on the following structure, which allows us to group sets of mutually non-laminar cycles into a single unit.

\begin{definition}
For $k\ge 4$, a \emph{$k$-frame $(a,b,P_1,..,P_k)$} is a structure of $\derivedgraph(G)$ in which:
\begin{itemize}
\item Both $a$ and $b$ are  vertices of  $\derivedgraph(G)$.
\item Each $P_i$ is an induced three-edge path from $a$ to $b$.
\item Each two distinct paths $P_i$ and $P_j$ are internally vertex disjoint.
\item The paths are numbered in their cyclic ordering (or its reverse) around $a$ and $b$ in the given planar embedding.
\item None of the paths $P_i$ contain one of the $v$--$f$--$v$ or $f$--$v$--$f$ triples that are forbidden in an allowable cycle.
\end{itemize}
The \emph{poles} of the frame are the vertices $a$ and $b$.
\end{definition}

The next two lemmas restrict the types of vertices in $\derivedgraph(G)$ that can be poles of frames.

\begin{figure}[t]
\centering\includegraphics[scale=0.5]{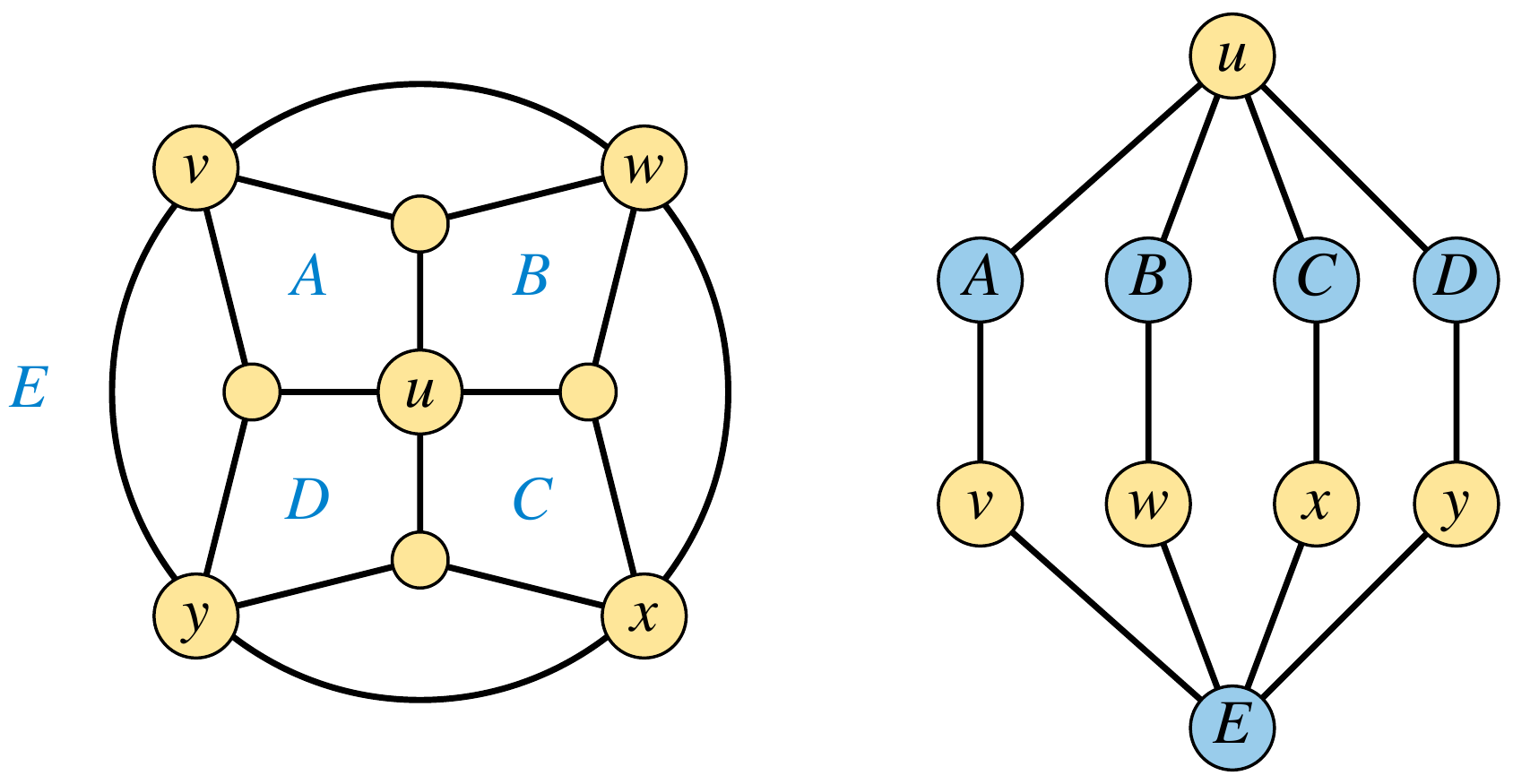}
\caption{A planar graph $G$ with nine vertices and nine faces (left), and an allowable $4$-frame $(u,E,uAvE,uBwE, uCxE, uDyE)$ in $\derivedgraph(G)$.}
\label{fig:frame}
\end{figure}

\begin{lemma}
\label{lem:pole-not-edge}
If $G$ is not $K_4$,
and $a$ is one pole of a frame in $\derivedgraph(G)$ then $a$ cannot correspond to an edge of~$G$.
\end{lemma}

\begin{proof}
Suppose for a contradiction that $a$ is the pole of a frame and corresponds to an edge $e$ of~$G$.
Then the frame can have only four paths, one for each of the four neighbours of $a$ in $\derivedgraph(G)$, alternating between vertices and faces of~$G$. The opposite pole $b$ must also correspond to an edge $f$ of $G$, because if $b$ corresponded to a face there would be no induced two-edge induced paths from the vertex neighbours of $a$ to it, and symmetrically if $b$ corresponded to a vertex there would be no two-edge induced paths from the face neighbours of $a$ to it. Symmetrically, the remaining four vertices of the four paths in the frame must be the four neighbours of $b$.

Thus, in $G$ this configuration can only arise when we have two edges $e$ and $f$ with the property that the endpoints of each edge lie on the faces adjacent to the other edge. For $G$ to avoid having two endpoints of these two edges as a $2$-cutset, it can only be $K_4$, with $e$ and $f$ as antipodal edges.
\end{proof}

\begin{lemma}
\label{lem:poles-vertex-face}
If $G$ is not $K_4$, then
for every frame in $\derivedgraph(G)$ the two poles correspond to one face and one vertex of~$G$.
\end{lemma}

\begin{proof}
By \autoref{lem:pole-not-edge}, each pole must correspond to a face or a vertex of~$G$. It is not possible for both poles to correspond to vertices of~$G$, because $\derivedgraph(G)$ does not contain any induced three-edge paths that start and end at vertices of~$G$. For the same reason it is not possible for both poles to correspond to faces of~$G$. Therefore, one pole must correspond to a face and the other to a vertex.
\end{proof}

\subsection{Frames from non-laminar cycles}

Frames capture the only ways in which allowable cycles can be non-laminar, as the next lemma shows.

\begin{lemma}
\label{lem:cross-makes-frame}
If  two allowable cycles cross in $\derivedgraph(G)$ then  their union is a
$4$-frame.
\end{lemma}

\begin{proof}
If two allowable cycles are non-laminar then their intersection is a subgraph with at least two connected components.
However, if the intersection contains two vertices that are one unit apart on one cycle, it contains the edge between them, because both cycles are induced. And if it contains two vertices that are two units apart on one cycle, then it contains the two-edge path connecting them, because (by \autoref{lem:unique-2-path}) there is only one way for an allowable cycle to connect these two vertices and the other cycle must connect them in the same way.
Therefore, in order for the intersection to contain two or more separate components, there must be exactly two components, each consisting of a single vertex, antipodal to each other. These two antipodal vertices form the poles of a $4$-frame.
\end{proof}

We can collect larger numbers of mutually crossing cycles into \emph{maximal frames}, defined below.

\begin{definition}
A $k$-frame is \emph{maximal} if there does not exist a $(k+1)$-frame with the same two poles and a superset of the same paths.
\end{definition}

When a frame is maximal, it encompasses all of the allowable paths between its poles:

\begin{lemma}
\label{lem:frames-contain-paths}
If $F$ is a maximal frame with poles $a$ and $b$, and $P$ is an allowable three-edge induced path from $a$ to $b$, then $P$ must be one of the paths of~$F$.
\end{lemma}

\begin{proof}
$P$ cannot be disjoint from all the paths, or else $F$ would not be maximal.
And an allowable induced path from $a$ to $b$ is determined by any one of its interior vertices, so $P$ cannot have a non-trivial intersection with a path in $F$ unless it coincides with that path.
Since it cannot be disjoint from all paths and it cannot have a partial intersection with any of the paths, it must be one of the paths.
\end{proof}

Because maximal frames cover all allowable paths, they also cover all of the allowable cycles through their two poles:

\begin{lemma}
\label{lem:avoid-poles}
If $C$ is an allowable cycle in a graph that is not $K_4$, $a$ and $b$ are vertices of $C$, and $F$ is a maximal frame with poles $a$ and $b$, then $C$ must be the union of two paths of $F$.
\end{lemma}

\begin{proof}
By \autoref{lem:poles-vertex-face} and the fact that the paths of a frame are induced, for $a$ and $b$ to be poles of $F$, they must be a non-incident vertex-face pair in $G$, which necessarily have distance at least three from each other in $\derivedgraph(G)$. Therefore, they can only be in antipodal positions of $C$, and $C$ must be formed by two induced paths of length three from $a$ to $b$. By \autoref{lem:frames-contain-paths}, these paths must both belong to the given maximal frame.
\end{proof}

\subsection{Big frames}

When a frame has sufficiently many paths, it cannot be crossed by any other allowable cycles than the ones formed by its paths.

\begin{definition}
A \emph{big frame} is a frame in $\derivedgraph(G)$ that has ten or more paths.
\end{definition}

\begin{lemma}
\label{lem:frame-laminarity}
Let $F$ be a maximal big frame. Then every allowable cycle of $\derivedgraph(G)$ is either formed by two paths of $F$, or lies between two paths of $F$ without crossing them.
\end{lemma}

\begin{proof}
Let $C$ be an allowable cycle that is not formed by two paths of $F$. Then by \autoref{lem:avoid-poles}, $C$ can only intersect $F$ at the two interior vertices of each path. It cannot cross a single path of $F$ twice, because the edge between the two crossing points would cause it to be non-induced. But the only other way for it to cross any of the paths of $F$ would be for it to loop around the two poles crossing each path exactly once, an impossibility because $C$ has only six vertices and there would necessarily be ten or more crossing points.
\end{proof}

It is not necessarily the case that every pair of allowable paths in a frame forms an allowable cycle. But when two allowable paths are far enough apart from each other in the cyclic ordering of paths around the poles of a frame, they do form an allowable cycle.

\begin{lemma}
\label{lem:frames-contain-allowable-cycles}
Let $F$ be a big frame and let $C$ be a cycle formed by paths that are separated from each other in both directions in $F$ by at least four other paths. Then $C$ is an allowable cycle. Every shift of $C$ is a canonical cycle that includes both poles of $F$.
\end{lemma}

\begin{proof}
Because of the separating paths, $C$ cannot be induced nor can it be the neighbourhood of a single vertex in $\derivedgraph(G)$. The separating paths also prevent $C$ from having non-allowable triples of vertices at the poles of $F$.
When $C$ is shifted, vertices of $C$ that correspond to edges of $G$ (incident to the face pole)
may shift to endpoints of those edges, belonging to adjacent paths of $F$. However, even after
this change, these endpoints will still be separated by two more of the four paths. Therefore,
they cannot be adjacent to each other in $G$ and it will not be possible to shift the face pole of $C$.
The resulting cycle is separating in $G$, again because of the two paths that remain, so it is canonical.
\end{proof}

\subsection{When there are no big frames}

Our eventual algorithm will subdivide $\derivedgraph(G)$ into subgraphs, some of which are big frames and others of which do not contain any big frames. As we show now, having no big frames implies that there are few allowable cycles and few conflicting pairs of allowable cycles, so the conflict graph will have linear size.

\begin{lemma}
\label{lem:diff-nbrs}
Suppose that $C$ is an allowable cycle in $\derivedgraph(G)$, $v$ is a vertex of $C$, and
$S$ is a set of allowable cycles that all cross $C$ at $v$. Then no two cycles in $S$ pass through the same two neighbours of $v$ in $\derivedgraph(G)$.
\end{lemma}

\begin{proof}
All cycles in $S$ must contain the vertex $w$ antipodal to $v$ in $C$, by \autoref{lem:cross-makes-frame}. And each cycle in $S$ is determined by $w$ and by the two neighbours of~$v$ that belong to the cycle, by \autoref{lem:unique-2-path}. Thus, there can be at most one cycle in $S$ for each pair of neighbours of~$v$.
\end{proof}

\begin{lemma}
\label{lem:many-cross-big-frame}
Suppose that $C$ is an allowable cycle in $\derivedgraph(G)$ and
$S$ is a set of $s$ allowable cycles that all cross $C$.
Then the union of $C$ and $S$ contains a $k$-frame for $k=\Omega(\sqrt s)$.
\end{lemma}

\begin{proof}
We may assume (with a factor of three loss in the size of $S$) that all cycles in $S$ cross $C$ at the same two antipodal points $v$ and $w$. By \autoref{lem:diff-nbrs} these cycles must have different pairs of neighbours at~$v$, and therefore there must be $\Omega(\sqrt s)$ neighbours among them, for otherwise there would not be enough pairs of neighbours to distinguish all the cycles in~$S$ from each other. By \autoref{lem:frames-contain-paths} each neighbour is represented by one of the paths in a frame that is maximal among the frames contained in the union of $C$ and~$S$.
\end{proof}

It follows from \autoref{lem:many-cross-big-frame} that, in a subgraph of $\derivedgraph(G)$ with no big frames, each allowable cycle crosses only a bounded number of others.
We will use this to show that, for certain subgraphs of $\derivedgraph(G)$, there are a linear number of canonical cycles and pairs of non-laminar canonical cycles, but we need to restrict the subgraphs in order to relate canonical cycles to allowable cycles in them.

\begin{definition}
An \emph{allowable subgraph} is a subgraph of $\derivedgraph(G)$ with the property that every canonical cycle in the subgraph is a shift of an allowable cycle that is also in the subgraph.
\end{definition}

We use the well-known crossing lemma of Ajtai et al.~\cite{AjtChvNew-TPC-82,Lei-CIV-83}
to transform a constant bound on the number of crossings per cycle into a linear bound on the total number of cycles.

\begin{lemma}[the crossing lemma~\cite{AjtChvNew-TPC-82,Lei-CIV-83}]
Let $G$ be a drawing in the plane of a simple graph (not a multigraph) with $n$ vertices and $m$ edges, with $m\ge 4n$,
with each intersection of two edges being either a shared endpoint of both edges or a proper crossing point. Then in the drawing, $\Omega(m^3/n^2)$ pairs of edges cross.
\end{lemma}

\begin{lemma}
\label{lem:few-allowable}
Let $H$ be a subgraph of $\derivedgraph(G)$ containing $h$ vertices and not containing any $k$-frame, where $h$ is variable but $k$ is an arbitrary fixed constant. Then $H$ contains $O(h)$ allowable cycles.
\end{lemma}

\begin{proof}
By \autoref{lem:diff-nbrs}, each pair of vertices in $H$ can be the antipodal vertices of $O(1)$
allowable cycles, for otherwise the cycles for some pair could be combined to form a large frame.

Consider choosing one allowable cycle through each antipodal pair, finding two representative points for the corresponding objects in a planar embedding of~$G$, and drawing one of the two paths between these antipodes in the chosen allowable cycle as a curve in $G$. By perturbing these curves so that they do not coincide in $G$ when the corresponding paths have a nonempty intersection in~$H$, it is possible to draw this system of curves in such a way that two curves cross if and only if they come from allowable cycles that cross. The system of curves has $h$ vertices and a number of curves proportional to the number of allowable cycles, and forms a drawing of a simple graph with these numbers of vertices and edges.

If there is to be no $k$-frame, then by \autoref{lem:many-cross-big-frame} each curve can be crossed only $O(1)$ times. And for this to happen, by the crossing lemma there can be only $O(h)$ curves.
\end{proof}

\begin{corollary}
Let $H$ be an allowable subgraph of $\derivedgraph(G)$ containing $h$ vertices and not containing any $k$-frame, where $h$ is variable but $k$ is an arbitrary fixed constant. Then the conflict graph $\conflictgraph(H)$ has $O(h)$ vertices and edges.
\end{corollary}

\subsection{Finding the frames}
 
\begin{lemma}
 \label{lem:find-big-frames}
The maximal big frames of $\derivedgraph(G)$ have total size that is linear in the size of $G$ and can be constructed from $G$ in time linear in the size of $G$.
\end{lemma}

\begin{proof}
The linear complexity of the set of maximal big frames follows from the fact that we can find a planar drawing of the graph that connects the pairs of poles of each maximal big frame, by routing the edge between each pair of poles along one of the paths of the frame: no two frames can have crossing edges, by the fact that any path of any big frame belongs to an allowable cycle (\autoref{lem:frames-contain-allowable-cycles}) and this cycle cannot cross any other frame (\autoref{lem:frame-laminarity}).

To construct the set of maximal big frames, we apply \autoref{lem:sgi2} to find all paths that belong to a big frame. We number the vertices of $\derivedgraph(G)$, label the paths by the two numbers at their endpoints, and radix sort the paths by these two-number labels. The sorted order groups together paths that have the same two endpoints, and the union of all paths with the same two endpoints is a maximal big frame.
\end{proof}

\section{Orienting the graph}

As part of our algorithms, it will be helpful to have a data structure that can distinguish one side of a cycle from another, or tell which of two cycles contains the other.
We will use this data structure for three purposes: to build a tree-decomposition from a set of laminar cycles, to choose cycles from frames in a way that allows each frame's cycles to be chosen independently without conflict from other frames, and to determine whether a $3$-cutset is $v$-non-shiftable.

\subsection{Counting inner faces}

We first describe a simple data structure for counting the inner faces of a cycle.

\begin{figure}
\centering\includegraphics[scale=0.44]{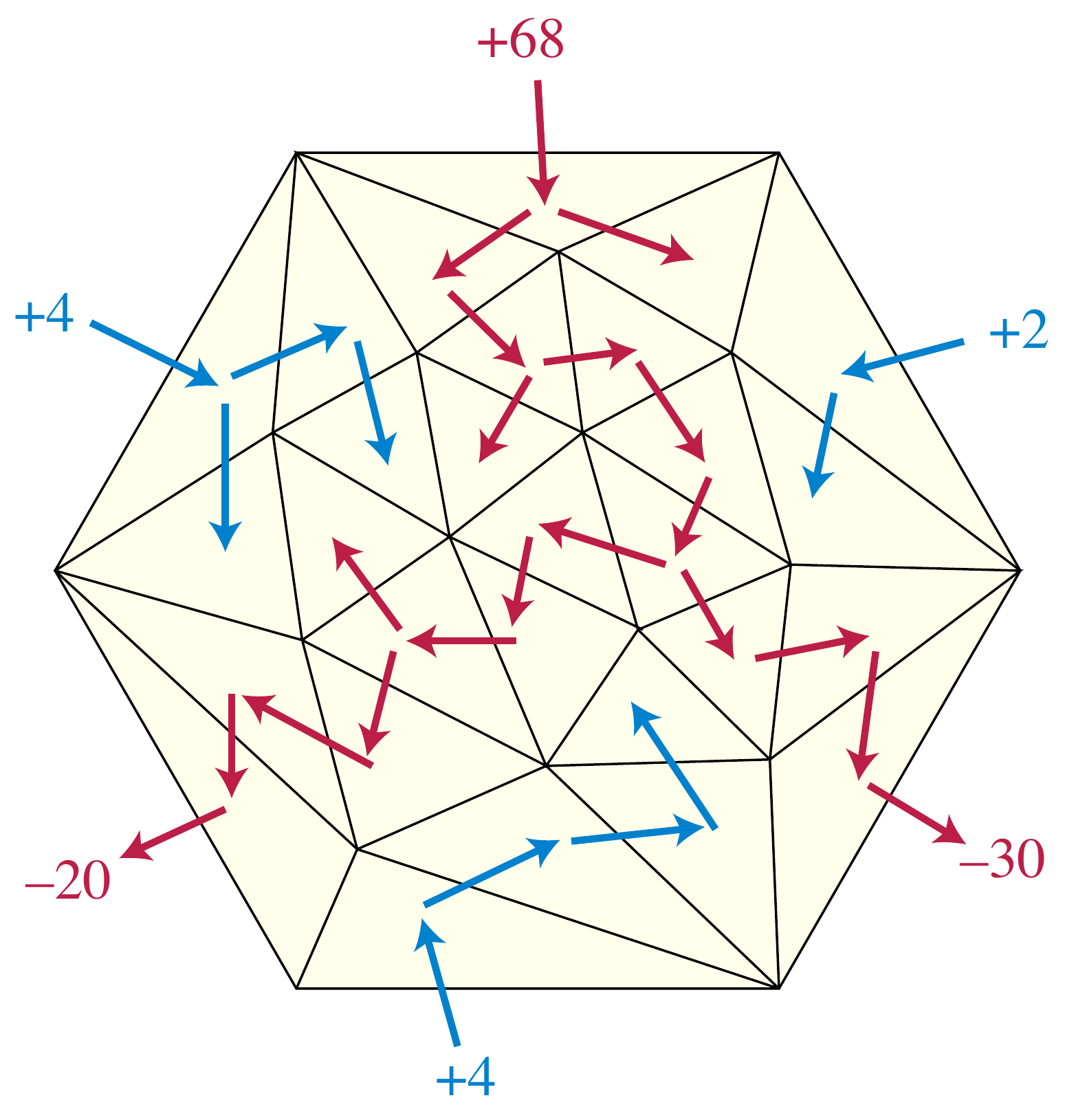}
\caption{Computing the number of faces inside a cycle as the signed sum of sizes of subtrees in a dual spanning tree. In this example, there are $28=68+2-30+4-20+4$ interior faces.}
\label{fig:subtree-sum}
\end{figure}

\begin{lemma}
\label{lem:facecount}
Let $G$ be an arbitrary embedded planar graph, with one of its faces $f$ chosen to be the outer face. Then in time linear in the size of $G$ we can construct a data structure such that, for any simple cycle $C$ in $G$, we can determine which side of $C$ is inside and
count the faces of $G$ inside $C$ in time $O(|C|)$.
\end{lemma}

\begin{proof}
Let $T$ be an arbitrary spanning tree of the dual graph of $G$, rooted at the dual vertex of $f$ and directed away from the root. Then we associate with each edge $e$ of $G$ two pieces of information: its orientation in $T$ (that is, which face of $G$ it is directed from and which it is directed to), and the number $t_e$ of dual vertices of $T$ in the subtree reached by that edge. These may be calculated by a straightforward bottom-up traversal of~$T$.

Then, to calculate the number of faces on one side of $C$, we compute $c=\sum_{e\in C}\pm t_e$,
where an edge $e$ contributes $+t_e$ to the sum if its dual edge in $T$ points into the chosen side of $C$, and $-t_e$ if its dual edge in $T$ points out from the chosen side (\autoref{fig:subtree-sum}).
If the chosen side is the inside of $C$, this sum will be positive, and will count the faces inside $C$.
If the chosen side is the outside, it will be negative, and $-c$ will count the faces inside~$C$.
\end{proof}

The same dual spanning tree can also be used to test containment of one cycle in another.

\begin{lemma}
\label{lem:facecont}
Let $G$ be an arbitrary embedded planar graph, with one of its faces $f$ chosen to be the outer face. Then in time linear in the size of $G$ we can construct a data structure such that, for any two simple cycles $C$ in $G$, we can determine whether $C$ contains $C'$ in time $O(|C|\cdot|C'|)$.
If we already know that $C$ and $C'$ are laminar, the time can be reduced to $O(|C|+|C'|)$.
\end{lemma}

\begin{proof}
We use the same spanning tree $T$ of the dual graph of $G$ used in \autoref{lem:facecount},
together with a standard trick for testing ancestry in trees:
We associate with each node of $T$ its index into the preorder and postorder traversals of the tree.
A node $x$ in $T$ is an ancestor of another node $y$ if and only if $x$ has both a lower preorder number and a higher postorder number than $y$.

To test whether cycle $C$ contains cycle $C'$, we determine which side of $C'$ is its inside (using \autoref{lem:facecount}) and then check, for each face $f$ adjacent to an edge of $C'$ on its inside, whether that face is also contained in $C$. If $C$ and $C'$ are known to be laminar, this check can be reduced to a single one of the faces adjacent to an edge of $C'$, together with a symmetric check for whether a single face adjacent to an edge of $C$ is contained in $C'$.

To check whether a face $f$ of $G$ is contained in a cycle $C'$, we check whether each edge of $C$ lies on the path from $f$ to the root of $T$. It is on this path if and only if the child endpoint of the edge is equal to $f$ or is an ancestor of $f$. If the number of edges of $C$ on this path is odd, $C$ contains $f$, and if this number is even then $C$ does not contain $f$.
\end{proof}

\subsection{Building a tree-decomposition}

\begin{lemma}
\label{lem:build-tree-decomposition}
Let $G$ be an embedded $3$-vertex-connected planar graph, and $S$ be a laminar set of canonical cycles in $\derivedgraph(G)$. 
Then in linear time we can construct a tree decomposition of adhesion three whose cuts (corresponding to the edges in the tree decomposition) are exactly the $3$-cutsets of $G$ that correspond to cycles in $S$.
\end{lemma}

\begin{proof}
Choose an outer face for $\derivedgraph(G)$ arbitrarily.
We apply \autoref{lem:facecount} to count the faces within each cycle of $S$ in linear time.
Next, we use a bucket sorting algorithm to sort $S$ by this number of faces, largest to smallest.
We make six copies of each cycle in this sorted list, one for each edge in each cycle.
We apply a stable bucket sorting algorithm, a second time, on the list of these copies, to sort them by their edge labels. The result of this second sorting algorithm can be partitioned into a collection of lists, one for each edge of $\derivedgraph(G)$,
of the cycles that use that edge in sorted order by the number of faces they contain. By applying \autoref{lem:facecont} we may then split this sorted list into at most two sorted lists of cycles that all contain each other, sorted from outer to inner.

Next, we construct the $3$-regular dual graph of $\derivedgraph(G)$ and remove from it the edges that belong to cycles in $S$. The connected components of this dual subgraph describe the pieces of $\derivedgraph(G)$ formed by cutting the plane along all cycle edges.

Next, we find a ``parent cycle'' for each cycle in $S$ and for each connected component of the dual subgraph. The parent cycle of a cycle $C$ or component $K$ is the innermost cycle that contains $C$ or $K$ (other than $C$ itself). To find this parent cycle, we apply the following rules:
\begin{itemize}
\item The parent cycle of the component of the dual subgraph containing the outer face is the outer face.
\item If a cycle $C$ in $S$ appears in one of the sorted lists of cycles that all contain each other, and is not the first cycle in that list, then its parent cycle is its predecessor in the list.
\item If a component $K$ has a boundary edge on one of its faces $f$ with a nonempty sorted list of cycles that all contain each other, and if the innermost cycle on this list contains $f$, then the parent cycle of $K$ is this innermost cycle.
\item If an edge of $\derivedgraph(G)$ has two sorted lists of cycles, then the outermost cycles of the two lists have the same parent cycle.
\item If An edge of $\derivedgraph(G)$ has one sorted list of cycles, and $f$ is the face incident to this edge that is not contained in any of these cycles, then the component containing $f$ and the outermost cycle of the list have the same parent cycle.
\end{itemize}
By applying a connected component algorithm to the undirected graph of ``same parent cycle'' relations described above, we can determine the parent cycle for every face or component in linear time.

We form a tree decomposition with one bag for the interior of each cycle of $S$, and one additional bag for the outside (the part of $G$ containing the chosen outer face).
Each vertex of $G$ that belongs to a cycle $C$ of $S$ is assigned both to the bag for $C$ and the bag for the parent cycle of $C$. A vertex may belong to multiple cycles, and be assigned in this way to multiple bags, but these bags will necessarily form a connected subtree rooted at the innermost cycle that contains but does not pass through the given vertex.
Each vertex of $G$ that does not belong to a cycle of $C$ is assigned to one bag, the bag for the parent cycle of the component of the dual subgraph that contains all the faces incident to that vertex.

The nesting structure of the cycles ensures that the bags and their parent relation form a tree.
Each edge in $G$ whose vertex in $\derivedgraph(G)$ belongs to a cycle in $S$ has its endpoints associated with the bag of that cycle, and each edge in $G$ that does not participate in $S$
has its endpoints associated with the bag that contains its component of the dual graph of $\derivedgraph(G)$, so in either case both endpoints are together in at least one bag.
Thus, the result is a valid tree-decomposition. The bag for a cycle $C$ and its parent share only three vertices in $G$, the ones in $C$, so the adhesion of the tree-decomposition is three and its cuts are the $3$-cutsets coming from cycles in $S$, as required.
\end{proof}

\subsection{Rooting the frames}

\begin{figure}[t]
\centering\includegraphics[scale=0.5]{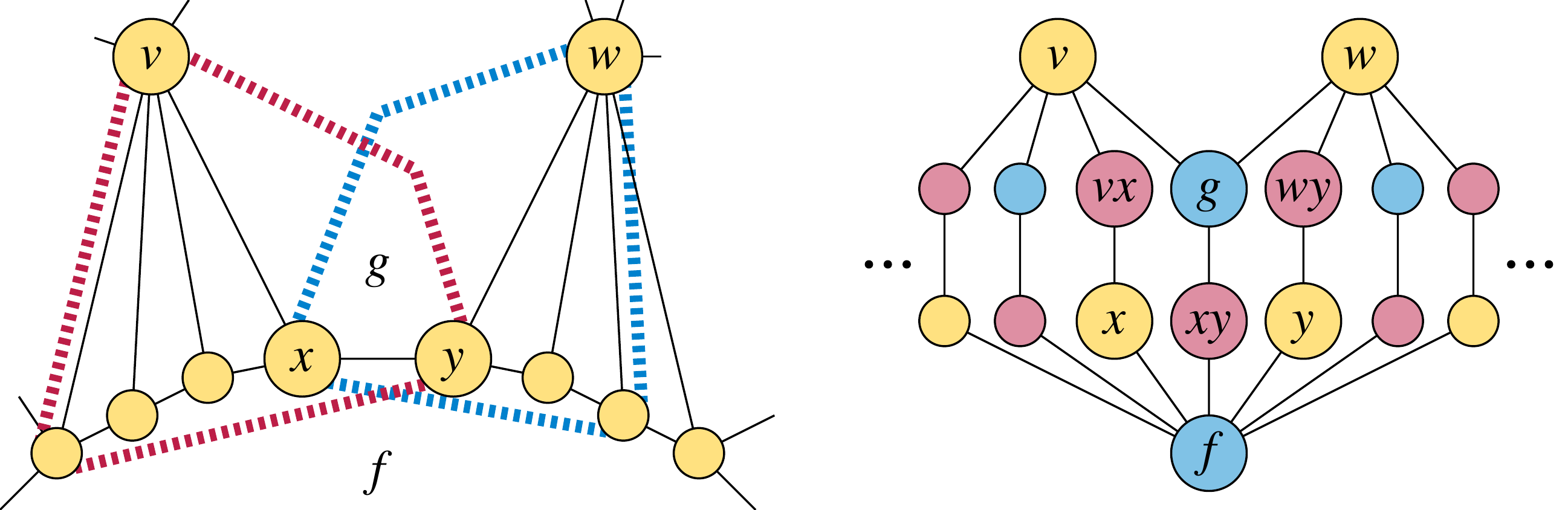}
\caption{A configuration in a graph $G$ (left) and two maximal frames in $\derivedgraph(G)$ (right) such that there is a non-laminar pair of canonical cycles shifted from an allowable cycle in each frame (red and blue dashed curves).}
\label{fig:overlapping-frames}
\end{figure}

The allowable cycles from two different maximal big frames are always laminar.
But if we choose an allowable cycle from two different maximal big frames,
and shift these cycles to become canonical cycles, they may become non-laminar (\autoref{fig:overlapping-frames}). We will use our data structure for determining the inside and outside of a cycle to help avoid this sort of conflict.

Again, let $f$ be (for now) an arbitrarily chosen outer face of $\derivedgraph(G)$.
We define the \emph{root cycle} of any frame to be the unique $6$-cycle composed of two paths of the frame that are outermost with respect to all the other $6$-cycles formed from other pairs of paths in the frame. Intuitively, the root cycle is composed of the two paths that are nearest $f$,
although this may not actually be true in terms of shortest path distance.

The following observation tells us that avoiding the root of each frame is sufficient
to allow us to choose laminar sets of cycles independently for all frames,
without having to worry whether the choice in one frame will affect what we do in a different frame.

\begin{observation}
\label{obs:safe-to-avoid-root}
Let $F_1$ and $F_2$ both be distinct big maximal frames in $\derivedgraph(G)$.
Choose an allowable cycle $C_i$ in each frame $F_i$ that is edge-disjoint from the root of its frame,
and shift both cycles to produce canonical cycles.
Then the two canonical cycles constructed in this way are necessarily laminar.
\end{observation}

When a frame is rooted, we may use its root to distinguish the two directions in which each of the paths of the frame may be shifted. Given an allowable cycle formed by two paths $P$ and $Q$ of the frame that are at least five steps apart from each other (so that by \autoref{lem:frames-contain-allowable-cycles} all their shifts are canonical), a shift may replace one of the paths, of the form vertex--face--edge--face (in the original graph $G$), by replacing the edge by one of its two endpoints (and then replacing vertex--face--vertex paths by vertex--edge--vertex paths when possible). Cycle $PQ$ separates these two endpoints, one of which is on the side of $PQ$ containing the root and the other of which is not. When we perform a shift that replaces an edge by an endpoint on the same side of $PQ$ as the root, we say that this shift is ``towards the root'' and when we replace an edge by an endpoint on the opposite side of $PQ$ from the root, we say that the shift is ``away from the root''.

We also need a similar laminarity result to \autoref{obs:safe-to-avoid-root} for cycles produced by shifting within a single frame.
We must be more careful here: even if we shift a single allowable cycle from a frame in two different ways, the two resulting canonical cycles may be non-laminar. For instance, this will happen if both paths of the allowable cycle can be shifted in two ways, and we choose the two shifts to be the ones that shift one path towards the root and the other path away from the root.
To prevent this non-laminarity, it suffices to constrain one path of the allowable cycle (or one shared path of two allowable cycles) to shift only in one of its two possible directions:

\begin{observation}
\label{obs:safe-to-shift-frame}
Let $P$, $Q$, and $R$ be three paths of the same big maximal frame in $\derivedgraph(G)$,
with $Q$ and $R$ not necessarily distinct from each other, but both at least five steps from $P$ in the cyclic order of paths of the frame. Form canonical cycles $C_1$ and $C_2$ by shifting $P\cup Q$ and $P\cup R$ respectively,
such that $P$ is shifted the same way in both $C_1$ and $C_2$ (but the shifts of the $Q$ and $R$ may be arbitrary). Then $C_1$ and $C_2$ either coincide or are laminar.
\end{observation}

\subsection{Testing non-shiftability}

\begin{lemma}
\label{lem:test-shiftable}
For a given $3$-connected planar graph $G$ and vertex $v$,
after linear time to build a data structure of linear size, we can test whether a given
canonical cycle $C$ in $\derivedgraph(G)$ corresponds to a $v$-non-shiftable $3$-cutset in constant time.
\end{lemma}

\begin{proof}
We choose a face of $\derivedgraph(G)$ incident to $v$ as the outer face, and build the data structure of \autoref{lem:facecount}.
Suppose that we wish to test whether a cycle $C$ is $v$-non-shiftable.
We can assume that $C$ does not contain $v$, because otherwise the answer is immediate.
For each vertex $w$ of $G$ that belongs to $C$, we first examine its neighbours in $C$ to determine whether they are edges to other vertices of $G$ in $C$.
This check gives us zero, one, or two neighbours of $w$ that are not on the $v$-side of $C$.
Next, we walk in $\derivedgraph(G)$ around the triangles surrounding $w$, starting from one of its two neighbours in $C$ and continuing  into the interior side of $C$, until either we find two edges of $G$ connecting $w$ to vertices interior to $C$, or we reach the other neighbour of $w$ in $C$ without finding enough neighbours.

If any $w$ does not have enough neighbours, we return that $C$ is not $v$-non-shiftable.
If we find enough neighbours for all three vertices, we return that $C$ is $v$-non-shiftable.
The test takes constant time per cycle $C$, because the walk for each vertex $w$ takes only a constant number of steps to terminate.
\end{proof}

\section{The main results}

We are ready to prove our main results, that maximal laminar sets of $3$-cutsets of various types can be found in linear time.

\subsection{Unrestricted $3$-cutsets}

\begin{theorem}
\label{thm:unrestricted}
Let $G$ be a $3$-connected planar graph. Then in time linear in the number of vertices of $G$, we can find a maximal laminar family of $3$-cutsets in~$G$, and a tree-decomposition of adhesion three whose cutsets are the cutsets in the maximal laminar family.
\end{theorem}

\begin{proof}
We perform the following steps.
\begin{enumerate}
\item Find a planar embedding of $G$, and construct $\derivedgraph(G)$ from it.
\item\label{step:choose-outer} Choose an arbitrary face of $\derivedgraph(G)$ to be the outer face, and initialize the face-counting data structure of \autoref{lem:facecount} and face-containment structure of \autoref{lem:facecont} on $\derivedgraph(G)$.
\item Apply \autoref{lem:find-big-frames} to find all maximal big frames in $\derivedgraph(G)$.
\item\label{step:framecycles} For each maximal big frame $F_i$, find a nearly-maximal laminar set of canonical cycles $S_i$, as follows.
\begin{enumerate}
\item Find the root cycle of frame $F_i$ by using the data structure of \autoref{lem:facecount}.
\item Let $P$ be a path of frame $F_i$ that is adjacent (in the cyclic order of the paths) to one of the two paths of the root cycle, but does not itself belong to the root cycle.
\item Let $A$ be the set of allowable cycles $A_j=P\cup Q_j$, where $Q_j$ is a path of frame $F_i$ that is at least five steps from $P$ in the cyclic order. For this choice of $Q_j$, it cannot be a path of the root cycle of $F_i$. Each $A_j$ is allowable by \autoref{lem:frames-contain-allowable-cycles}.
\item For each allowable cycle $A_j$, construct one or two canonical cycles $C_j$, $C_j'$
by shifting $A_j$ away from the root on path $P$, and in either of the two possible shift directions on path $Q_j$. (The third vertex of $G$ on the cycle, at the point where all the paths of the frame meet, cannot be shifted.)
\item Let $S_i$ be the set of all canonical cycles $C_j$, $C_j'$ constructed in this way.
\end{enumerate}
The sense in which $S_i$ is nearly-maximal is that the only possible canonical cycles that are not in but are laminar with $S_i$, and are shifts of allowable cycles formed by two paths of $F_i$, are in the two regions between $P$ and the paths three steps away. Within each of these regions, there are too few paths of $F$ to form a big frame.
\item Let $S=\cup S_i$, the set of all canonical cycles found for all maximal big frames (removing any duplicate copies of cycles found in this way). By \autoref{obs:safe-to-avoid-root} and \autoref{obs:safe-to-shift-frame}, $S$ is laminar.
\item Cut $\derivedgraph(G)$ along all edges in cycles of $S$, producing a single disconnected graph $H$, embedded onto a surface with boundary cycles of the graph along all of the cuts, and with the same set of faces as $\derivedgraph(G)$. To do so, form the dual graph of $\derivedgraph(G)$, remove from it the dual edges corresponding to edges in cycles of $S$, and construct the connected components of the resulting dual subgraph. Within each connected component, add back the copies of the edges and vertices of $\derivedgraph(G)$ surrounding each face of $\derivedgraph(G)$ whose dual vertex belongs to the component. (Using different copies of these edges and vertices for each different component.) Because we cut along nearly-maximal laminar sets of cycles for all big frames, $H$ has no big frames.
\item\label{step:solve-frameless} Apply \autoref{lem:solve-frameless} to find a maximal laminar set $R$ of canonical cycles within $H$. The $3$-cutsets for cycles in $R\cup S$ will be our maximal laminar family.
\item Apply \autoref{lem:build-tree-decomposition} to build and return a tree-decomposition for this set of $3$-cutsets. 
\end{enumerate}
\end{proof}

\subsection{Nontrivial $3$-cutsets}

Recall that a $3$-cutset is nontrivial when it does not consist of the three neighbours of a single vertex. The canonical cycles formed by paths five or more steps apart in a frame automatically produce nontrivial $3$-cutsets, because the two or more remaining paths of the frame inside or outside the canonical cycle contain too many vertices of $\derivedgraph(G)$ to correspond to a single vertex of~$G$. Therefore, the only step of the algorithm of \autoref{thm:unrestricted} that needs to be modified, to produce laminar sets of nontrivial $3$-cutsets, is Step~\ref{step:solve-frameless}, in which we use \autoref{lem:solve-frameless} to find canonical cycles that are not associated with big frames.

\begin{theorem}
\label{thm:nontrivial}
Let $G$ be a $3$-connected planar graph. Then in time linear in the number of vertices of $G$, we can find a maximal laminar family of nontrivial $3$-cutsets in~$G$, and a tree-decomposition of adhesion three whose cutsets are the cutsets in the maximal laminar family.
\end{theorem}

\begin{proof}
We modify Step~\ref{step:solve-frameless} of the algorithm of \autoref{thm:unrestricted} so that,
after constructing $\conflictgraph(H)$ and before finding a maximal independent set in it,
we check whether each vertex of $\conflictgraph(H)$ represents a trivial or nontrivial $3$-cutset, and we delete the vertices of $\conflictgraph(H)$ that represent trivial cutsets before computing a maximal independent set in the remaining subgraph of~$\conflictgraph(H)$.
\end{proof}

\subsection{Non-shiftable $3$-cutsets}

To apply the same method to find a laminar set of $v$-non-shiftable $3$-cutsets, we need to avoid shiftable cutsets that come from the frames as well as the ones remaining after we cut the frame cycles. The following lemma shows that there will still remain enough $3$-cutsets for us to perform the cutting procedure in almost the same way, without leaving any uncut big frames.

\begin{lemma}
\label{lem:enough-non-shiftable}
Let $v$ be any vertex of a $3$-connected planar graph $G$, and select as the outside face of $\derivedgraph(G)$ any of the faces incident to~$v$. Let $F$ be a big frame of $\derivedgraph(G)$,
and let $A$ be an allowable cycle formed by two paths of $F$, at least five steps from each other in both directions around the frame, with neither path being part of the root cycle (relative to the choice of outer face). Let $C$ be a canonical cycle formed by shifting $A$ away from $v$. Then $C$ is $v$-non-shiftable.
\end{lemma}

\begin{proof}
Let $u$ and $f$ be the two poles of $F$, a vertex and face of $G$ respectively.
If $u$ is adjacent to an edge in $A$, then the other endpoint of the edge is on or inside the $3$-cutset of $C$, and if $u$ is instead adjacent to a face in $A$, then one of its neighbours (in $G$) on the boundary of that face is on or inside the cutset. Because the paths forming $A$ are far enough apart, $u$ has two distinct neighbours on or inside the cutset, as is required for a non-shiftable cycle.

Now let $w$ be any other vertex in the $3$-cutset of $C$.
Then $w$ has at least one neighbour that is on or inside the cutset, adjacent to it (in $G$) around the boundary of $f$. If $w$ is adjacent in $G$ to $u$, it has two such neighbours.
The only remaining possibility is that allowable cycle $A$ connects $w$ to $u$ via another face $g$ of $G$, distinct from~$f$. It is possible for faces $f$ and $g$ to share an edge, but if they do
then $w$ must be the inner endpoint of the edge with respect to cycle $A$ and the choice of outer face, because of the way we chose $C$ to be the shift of $A$ away from~$v$.
Therefore, the inner neighbour of $w$ on $g$ is distinct from the inner neighbour of $w$ on $f$, and $w$ has two distinct neighbours on or inside the cutset.

We have shown that all vertices of the cutset of $C$ have two neighbours on or inside the cutset.
Therefore, $C$ satisfies the definition of $v$-non-shiftability.
\end{proof}

\begin{theorem}
\label{thm:non-shift}
Let $G$ be a $3$-connected planar graph, and let $v$ be any vertex of~$G$. Then in time linear in the number of vertices of $G$, we can find a maximal laminar family of $v$-non-shiftable $3$-cutsets in~$G$, and a tree-decomposition of adhesion three whose cutsets are the cutsets in the maximal laminar family.
\end{theorem}

\begin{proof}
We modify Step~\ref{step:choose-outer} of the algorithm of \autoref{thm:unrestricted}
so that it chooses a face of $\derivedgraph(G)$ that is incident to~$v$ as the outer face.
We modify Step~\ref{step:framecycles} of the algorithm
so that it tests each of the canonical cycles it generates for $v$-non-shiftability using \autoref{lem:test-shiftable} and keeps only the ones that pass this test.
By \autoref{lem:enough-non-shiftable} there will still be enough remaining non-shiftable canonical cycles to cut the same set of frame paths as before.

Then, we modify Step~\ref{step:solve-frameless} of the algorithm so that,
after constructing $\conflictgraph(H)$ and before finding a maximal independent set in it,
we check whether each vertex of $\conflictgraph(H)$ represents a $v$-non-shiftable $3$-cutset, and we delete the vertices of $\conflictgraph(H)$ that represent shiftable cutsets before computing a maximal independent set in the remaining subgraph of~$\conflictgraph(H)$.
\end{proof}

\newcommand{\ISBN}[1]{ISBN #1}
\bibliographystyle{amsplainurl}
\bibliography{planar3cut}

\providecommand{\bysame}{\leavevmode\hbox to3em{\hrulefill}\thinspace}
\providecommand{\MR}{\relax\ifhmode\unskip\space\fi MR }
\providecommand{\MRhref}[2]{%
  \href{http://www.ams.org/mathscinet-getitem?mr=#1}{#2}%
}
\providecommand{\href}[2]{#2}
\begin{thebibliography}{10}

\bibitem{AjtChvNew-TPC-82}
M.~Ajtai, V.~Chv{\'a}tal, M.~M. Newborn, and E.~Szemer{\'e}di,
  \emph{{Crossing-free subgraphs}}, Theory and practice of combinatorics,
  North-Holland Math. Stud., vol.~60, North-Holland, Amsterdam, 1982,
  pp.~9{--}12, \MR{806962}.

\bibitem{ChaEpp-JGAA-13}
E.~Chambers and D.~Eppstein, \emph{{Flows in one-crossing-minor-free graphs}},
  J. Graph Algorithms {\&} Applications \textbf{17} (2013), no.~3, 201{--}220,
  \href {http://dx.doi.org/10.7155/jgaa.00291} {\path{doi:10.7155/jgaa.00291}},
  \MR{3043209}.

\bibitem{ChiNis-SJC-85}
N.~Chiba and T.~Nishizeki, \emph{{Arboricity and subgraph listing algorithms}},
  SIAM Journal on Computing \textbf{14} (1985), no.~1, 210{--}223, \href
  {http://dx.doi.org/10.1137/0214017} {\path{doi:10.1137/0214017}},
  \MR{774940}.

\bibitem{DiBTam-ICALP-90}
G.~Di~Battista and R.~Tamassia, \emph{{On-line graph algorithms with
  SPQR-trees}}, Proc. 17th Internat. Colloq. Automata, Languages and
  Programming (ICALP 1990), Lect. Notes in Comput. Sci., vol. 443, Springer,
  1990, pp.~598{--}611, \href {http://dx.doi.org/10.1007/BFb0032061}
  {\path{doi:10.1007/BFb0032061}}.

\bibitem{Dor-STACS-10}
F.~Dorn, \emph{{Planar subgraph isomorphism revisited}}, 27th Int. Symp. Theor.
  Aspects. of Comp. Sci. (STACS 2010), Leibniz Int. Proc. Inform. (LIPIcs),
  vol.~5, Schloss Dagstuhl{--}Leibniz-Zent. Inform., Dagstuhl, Germany, 2010,
  pp.~263{--}274, \href {http://dx.doi.org/10.4230/LIPIcs.STACS.2010.2460}
  {\path{doi:10.4230/LIPIcs.STACS.2010.2460}}, \MR{2853927}.

\bibitem{Epp-JGAA-99}
D.~Eppstein, \emph{{Subgraph isomorphism in planar graphs and related
  problems}}, J. Graph Algorithms Appl. \textbf{3} (1999), no.~3, 1{--}27,
  \href {http://dx.doi.org/10.7155/jgaa.00014} {\path{doi:10.7155/jgaa.00014}},
  \MR{1750082}.

\bibitem{EppVaz-18}
D.~Eppstein and V.~V. Vazirani, \emph{{NC algorithms for perfect matching and
  maximum flow in one-crossing-minor-free graphs}}, preprint, 2018, \href
  {http://arxiv.org/abs/1802.00084} {\path{arXiv:1802.00084}}.

\bibitem{GutMut-GD-01}
C.~Gutwenger and P.~Mutzel, \emph{{A linear time implementation of
  SPQR-trees}}, Proc. 8th Int. Symp. Graph Drawing (GD 2000), Lect. Notes in
  Comput. Sci., vol. 1984, Springer, 2001, pp.~77{--}90, \href
  {http://dx.doi.org/10.1007/3-540-44541-2_8}
  {\path{doi:10.1007/3-540-44541-2_8}}.

\bibitem{HopTar-CACM-73}
J.~Hopcroft and R.~Tarjan, \emph{{Algorithm 447: efficient algorithms for graph
  manipulation}}, Commun. ACM \textbf{16} (1973), no.~6, 372{--}378, \href
  {http://dx.doi.org/10.1145/362248.362272} {\path{doi:10.1145/362248.362272}}.

\bibitem{KawLiRee-15}
K.~Kawarabayashi, Z.~Li, and B.~Reed, \emph{{Connectivity preserving iterative
  compaction and finding 2 disjoint rooted paths in linear time}}, Electronic
  preprint arxiv:1509.07680, 2015.

\bibitem{Lei-CIV-83}
T.~Leighton, \emph{{Complexity Issues in VLSI: Optimal Layouts for the
  Shuffle-Exchange Graph and Other Networks}}, Foundations of Computing Series,
  MIT Press, Cambridge, MA, 1983.

\bibitem{Mac-DMJ-37}
S.~Mac~Lane, \emph{{A structural characterization of planar combinatorial
  graphs}}, Duke Math. J. \textbf{3} (1937), no.~3, 460{--}472, \href
  {http://dx.doi.org/10.1215/S0012-7094-37-00336-3}
  {\path{doi:10.1215/S0012-7094-37-00336-3}}, \MR{1546002}.

\bibitem{NesOss-S-12}
J.~Ne{\v{s}}et{\v{r}}il and P.~Ossona~de Mendez, \emph{{Sparsity: Graphs,
  Structures, and Algorithms}}, Algorithms and Combinatorics, vol.~28,
  Springer, 2012, \href {http://dx.doi.org/10.1007/978-3-642-27875-4}
  {\path{doi:10.1007/978-3-642-27875-4}}, \MR{2920058},
  \ISBN{978-3-642-27874-7}.

\bibitem{StrThiWag-TCS-16}
Simon Straub, Thomas Thierauf, and Fabian Wagner, \emph{{Counting the number of
  perfect matchings in $K_5$-free graphs}}, Theory of Computing Systems
  \textbf{59} (2016), no.~3, 416{--}439, \href
  {http://dx.doi.org/10.1007/s00224-015-9645-1}
  {\path{doi:10.1007/s00224-015-9645-1}}, \MR{3539928}.

\bibitem{ThiWag-CJTCS-14}
Thomas Thierauf and Fabian Wagner, \emph{{Reachability in $K_{3,3}$-free and
  $K_5$-free graphs is in unambiguous logspace}}, Chicago Journal of
  Theoretical Computer Science (2014), A2:1{--}A2:29, available from
  \url{https://cjtcs.cs.uchicago.edu/articles/2014/2/contents.html},
  \MR{3201618}.

\bibitem{Wag-MA-37}
K.~Wagner, \emph{{{\"U}ber eine Eigenschaft der ebenen Komplexe}}, Math. Ann.
  \textbf{114} (1937), 570{--}590, \href {http://dx.doi.org/10.1007/BF01594196}
  {\path{doi:10.1007/BF01594196}}.

\end{thebibliography}

\end{document}